%% file: ms.tex
\documentclass[journal=jacsat,manuscript=article]{achemso}

\usepackage[T1]{fontenc} 
\usepackage{hyperref}
\usepackage{amsfonts,amsmath,amssymb,amsthm}
\usepackage{bm,nicefrac}
\usepackage{longtable, booktabs,array}
\usepackage{makecell}
\usepackage{float}
\usepackage[utf8]{inputenc}
\usepackage{xr}
\usepackage{graphicx}
\DeclareUnicodeCharacter{0301}{\'{e}}
\DeclareUnicodeCharacter{2009}{\,}
\DeclareUnicodeCharacter{03BA}{$\kappa$}
\DeclareUnicodeCharacter{03B2}{$\beta$}
\DeclareUnicodeCharacter{2212}{-}

\SectionNumbersOn

\makeatletter
\def\maxwidth{\ifdim\Gin@nat@width>\linewidth\linewidth\else\Gin@nat@width\fi}
\def\maxheight{\ifdim\Gin@nat@height>\textheight\textheight\else\Gin@nat@height\fi}
\makeatother
\setkeys{Gin}{width=\maxwidth,height=\maxheight,keepaspectratio}

\title{Synthesis-driven design of 3D molecules for structure-based drug discovery using geometric transformers}

\makeatletter
\newcommand*{\addFileDependency}[1]{
  \typeout{(#1)}
  \@addtofilelist{#1}
  \IfFileExists{#1}{}{\typeout{No file #1.}}
}
\makeatother

\newcommand*{\myexternaldocument}[1]{%
    \externaldocument{#1}%
    \addFileDependency{#1.tex}%
    \addFileDependency{#1.aux}%
}

\myexternaldocument{si-template}

\author{Yibo Li}
\affiliation{
  Center for Life Sciences, Academy for Advanced Interdisciplinary Studies,
  Peking University, Beijing 100871, China}
\author{Jianfeng Pei}
\affiliation{
  Center for Quantitative Biology, Academy for Advanced Interdisciplinary
  Studies, Peking University, Beijing 100871, China}
\email{jfpei@pku.edu.cn}
\author{Luhua Lai}
\affiliation{
  Center for Life Sciences, Academy for Advanced Interdisciplinary Studies,
  Peking University, Beijing 100871, China}
\alsoaffiliation{
  Center for Quantitative Biology, Academy for Advanced Interdisciplinary
  Studies, Peking University, Beijing 100871, China}
\altaffiliation{
  BNLMS, College of Chemistry and Molecular Engineering, Peking University,
  Beijing 100871, China
}
\email{lhlai@pku.edu.cn}

\begin{document}

  \begin{abstract}
    \input{abstract.tex}

  \end{abstract}


  \input{article.tex}

  \bibliography{main.bib}{}

\end{document}



  \input{si.tex}

  \bibliography{main.bib}{}

%% file: abstract.tex
Finding drug-like compounds with high bioactivity is essential for drug
discovery, but the task is complicated by the high cost of chemical
synthesis and validation. With their outstanding performance in \emph{de
novo} drug design, deep generative models represent promising tools for
tackling this challenge. In recently years, 3D molecule generative
models have gained increasing attention due to their ability to directly
utilize the 3D interaction information between the target and ligand.
However, it remains challenging to synthesize the molecules generated by
these models, limiting the speed of bioactivity validation and further
structure optimization. In this work, we propose DeepLigBuilder+, a deep
generative model for 3D molecules that combines structure-based \emph{de
novo} drug design with a reaction-based generation framework. Besides
producing 3D molecular structures, the model also proposes synthetic
pathways for generated molecules, which greatly assists the
retro-synthetic analysis. To achieve this, we developed a new way to
enforce the synthesizability constraint using a tree-based organization
of purchasable building blocks. This method enjoys high scalability and
is compatible with existing atom-based generative models. Additionally,
for structure-based design tasks, we developed an SE(3)-equivariant
transformer conditioned on the shape and pharmacophore-based inputs, and
combine it with the Monte Carlo tree search. Using the ATP-binding
pocket of BTK and the NAD+ binding pocket of PHGDH for case studies, we
demonstrate that DeepLigBuilder+ is capable of enriching drug-like
molecules with high predicted binding affinity and desirable interaction
modes while maintaining the synthesizability constraint. We believe that
DeepLigBuilder+ is a powerful tool for accelerating the process of drug
discovery, and represents an important step towards a fully automated
design-synthesis-evaluation workflow for molecule design.

%% file: article.tex
\section{Introduction}

The high financial cost and low success rate of drug discovery place
tremendous challenges in finding treatments for important
diseases\cite{Paul.2010}. To address those challenges, computational
methods have been developed to find promising compounds from the vast
space of chemical structures for subsequent biological validation.
Computational virtual screening (VS) have been widely used, which
filters chemical libraries using scoring functions\cite{Trott.2010}
for favorable compounds. In spite of their success in finding bioactive
molecules\cite{Lyu.2019}, VS is constrained by the screening library
it uses. Small libraries may have limited coverage of the chemical
space, and large ones impose high computational costs for the screening
process and require specialized software and hardware
platforms\cite{Gorgulla.2020}. \emph{De novo} drug design, which
uses computational algorithms to generate molecule structures from
scratch\cite{Schneider.2005}, provides an option to explore new
chemical space beyond libraries of existing compounds.

Over the decades, a variety of \emph{de novo} drug design programs have
been proposed, such as LEGEND\cite{Nishibata.1991},
LUDI\cite{Bohm.1992}, CONCEPTS\cite{Pearlman.1993} and
LigBuilder\cite{Wang.2000,Yuan.2011,Yuan.2020}, many of which have
been used to design bioactive molecules with successful experimental
validation
\cite{Schneider.2012,Schneider.2019,Ni.2009,Shang.2014,Park.2013}.

In recent years, deep molecule generative models have emerged as a new
class of promising methods for \emph{de novo} drug
design\cite{Xu.2019}. Using deep learning, models can automatically
learn traits of desirable molecule structures from the training data,
with little need for manual intervention. This contrasts significantly
with traditional \emph{de novo} design programs, which in general
require extensive efforts to design the search rules and scoring
functions. The advantages of deep generative models have helped to spawn
a series of research aiming to utilize them for drug discovery. Those
works range from exploring different molecule representations, including
SMILES\cite{Segler.2017} and molecular
graph\cite{Li.2018,Jin.2018,You.2018}, to testing with various
training methods, such as VAE\cite{Gomez-Bombarelli.2018},
GAN\cite{Guimaraes.2017,Cao.2018} and
RL\cite{Olivecrona.2017,You.2018lit}. As a result, a wide range of
models has been proposed to address various issues related to drug
design based on molecular properties\cite{Gomez-Bombarelli.2018},
pharmacophores\cite{Imrie.2020}, scaffolds\cite{Li.2019} and
targets\cite{Zhavoronkov.2019}.

Most deep generative models for molecules have been focused on designing
the 2D (topological) chemical structures, but the foundation of
bioactivity lies in the interactions between the 3D structures of
targets and ligands. Directly generating 3D molecules inside the target
binding pocket can help the model to better utilize the interaction
information. Additionally, it can reduce the need for ligand-based
information, potentially leading to molecules with higher novelty. Those
benefits have led to growing attention in developing 3D generative
models of molecules based on target information. Earlier approaches
include models that convert 3D pocket information into SMILES strings,
such as LiGANN\cite{Skalic.2019ncg} and the pocket-conditioned RNN
proposed by Xu et al.\cite{Xu.2021}, but the generated structure
only contains topological information. In order to directly generate 3D
structures, Masuda et al. \cite{Masuda.2020} proposed liGAN, which
uses VAE based on 3D-CNN to generate atomic density grids, and later
convert the grids to 3D molecules. To avoid the conversion between
different representations, we previously proposed
DeepLigBuilder\cite{Li.2021} to directly produce 3D molecules inside
pockets using graph generative models and Monte Carlo tree search. Other
graph generative models, such as Pocket2Mol\cite{Peng.2022}, use
equivariant networks to encode pocket information as conditional inputs.
More recent works have experimented with diffusion models for 3D
molecule generation, such as DiffLinker\cite{Igashov.2022}, which
features a permutation invariant way of generation compared to
autoregressive models.

Despite progress in structure-based 3D deep generative models, there is
still a critical issue to be addressed: the synthesizability of
generated molecules. Chemical synthesis is a common rate-limiting step
in medicinal chemistry research. Since \emph{de novo} design programs
are not constrained by any compound library, it is easier for these
methods to propose molecules that are challenging to synthesize,
especially in objective-directed situations\cite{Gao.2020pqg},
making experimental validation difficult. A solution to this problem is
to use synthetically aware models\cite{Coley.2020}. Those models
generate molecules by generating their synthetic path, using explicit
building blocks and chemical reactions. Such approaches have been
relatively common in traditional \emph{de novo} design programs, such as
SYNOPSIS\cite{Vinkers.2003} and DOGS\cite{Hartenfeller.2012},
but is largely absent from early deep learning methods. More recently,
an increasing number of models have been proposed to integrate this
approach with deep generative networks, including
MoleculeChef\cite{Bradshaw.2019}, DoG-AE and
DoG-Gen\cite{bradshaw2020barking}, PGFS\cite{Gottipati.2020} and
SynNet\cite{Gao.2021}. Those methods have shown promising results,
but they are largely focused on 2D molecule design, which, as discussed
before, inherits several limitations compared to recent 3D generative
models. Based on the discussions above, we believe that it is highly
beneficial to develop a deep generative model that can perform
pocket-based 3D molecule design while ensuring the synthesizability of
the generated molecules. In this work, we combine geometric deep
learning and synthesizability constraints to develop a new \emph{de
novo} drug design program, DeepLigBuilder+. The program follows a
reaction-based scheme for generating drug-like molecules, while at the
same time produces their 3D conformations, making it easy to be applied
to structure-based design tasks. Specifically, we use a transformer
network to generate 3D molecular graphs atom-by-atom, while at each
step, we mask inappropriate atom and bond types from the action space so
that the output structure is guaranteed to be inside the user-provided
reactant dataset (represented as synthons). In order to incorporate 3D
pocket information, we trained an SE(3)-equivariant transformer network
conditioned on pharmacophore and shape information, and combine it with
a reinforcement learning module based on Monte Carlo tree search (MCTS).
To demonstrate the capability of DeepLigBuilder+ in drug design
applications, we use it to design inhibitors targeting the ATP-binding
pocket of Bruton's tyrosine kinase (BTK), as well as the
NAD\(^+\)-binding pocket of human phosphoglycerate dehydrogenase
(PHGDH). In both cases, DeepLigBuilder+ generated molecules with high
predicted binding affinity and favorable binding modes while maintaining
the enforced synthesizability constraint.

\begin{figure}
\hypertarget{fig:network}{%
\centering
\includegraphics[width=0.8\textwidth,height=\textheight]{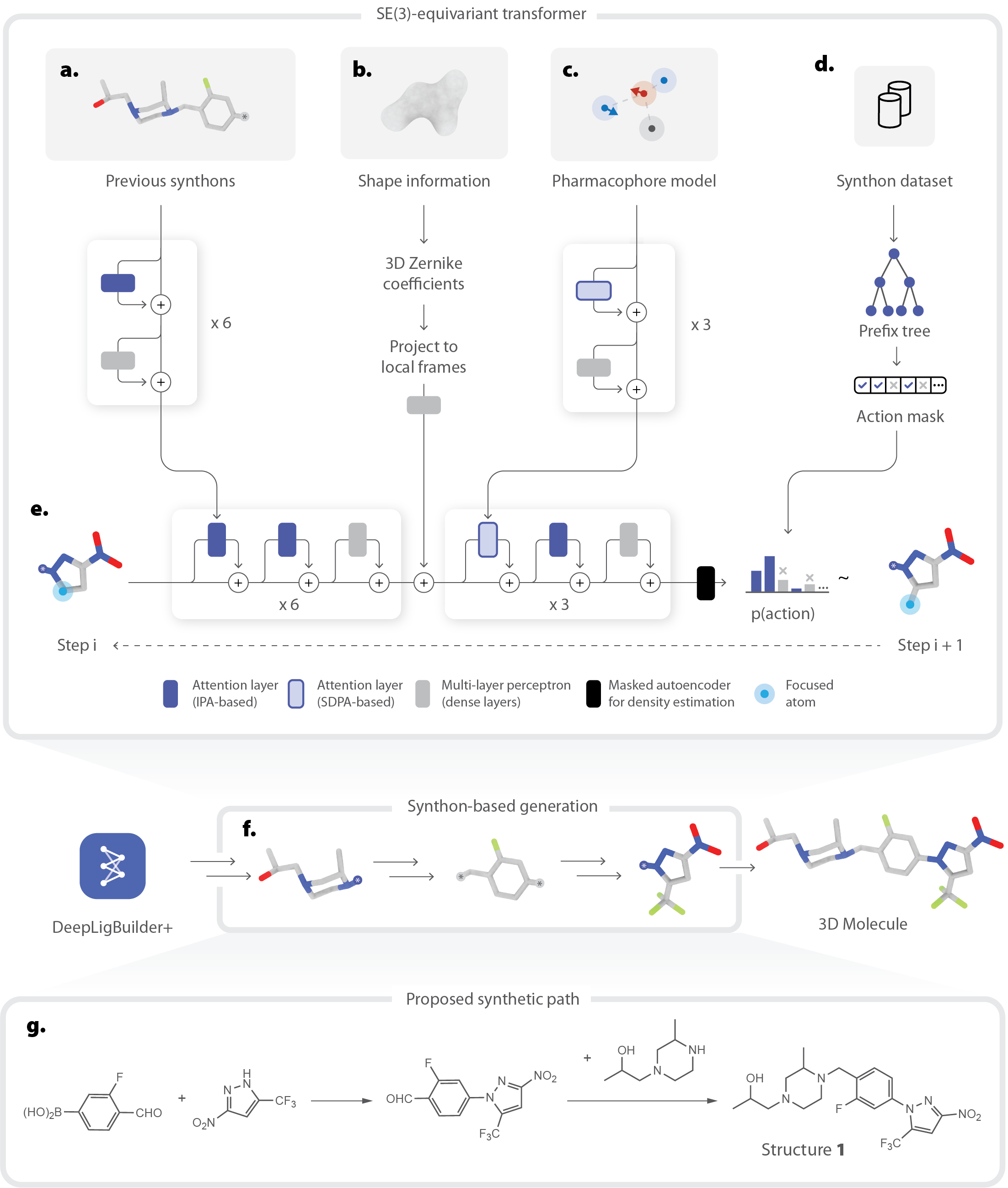}
\caption{An overview of DeepLigBuilder+. It generates synthesizable 3D
molecules following a reaction-based method using synthons (\textbf{f}).
When producing each synthon, the model adopts a graph-based generation
scheme (\textbf{e}) that iteratively edits the molecular graph by adding
new nodes (atoms) or edges (bonds). The decisions of how to perform
those edits are made by a transformer network (\textbf{a-e}). The
encoders are used to process the input information, such as previously
generated synthons (\textbf{a}). For structure-based generation tasks,
the encoder also receives shape(\textbf{b}) and
pharmacophore(\textbf{c})-based inputs. The decoder (\textbf{e}) uses
those input information to produce a state embedding, which is later
used by the policy network (based on MADE blocks) to output a
distribution in the action space. We apply action masks at each step to
constrain the generation trajectory so that it only produces synthons
that can be converted into purchasable building blocks (\textbf{d}).
Finally, the generated synthons can be converted to a synthetic route,
with explicit reactants and reaction types
(\textbf{g}).}\label{fig:network}
}
\end{figure}

\section{Methods}

In this section, we give a brief account of the architecture of
DeepLigBuilder+. The implementation details for DeepLigBuilder+ are
provided in the Supplementary Methods (Section S\ref{sec:s-method}).

To ensure high synthetic accessibility, DeepLigBuilder+ generates
molecules one reactant at a time and then produces the resulting
molecules using corresponding reaction rules. Unlike previous methods,
DeepLigBuilder+ also generates 3D conformation of the molecule for
subsequent structure-based design tasks. Since many reactions involve
large conformational changes, instead of directly generating 3D
structures of reactants, DeepLigBuilder+ first generates synthon
structures and later covert them to the corresponding reactants, as
shown in Figure \ref{fig:network}\textbf{a} and \textbf{g}. Synthons are
hypothetical reactants that have one or more open valences with specific
reactivity. A reactant can be converted to a synthon by extracting the
substructure of the product that is derived from this reactant. In this
way, adding a new synthon to the molecule will not affect the
conformation of previous synthons, making it more suitable for 3D
generation tasks. In this work, we use the global stock of Enamine
building blocks as the reactant set and use the reactions collected by
Hartenfeller et al.\cite{Hartenfeller.2011} . Each product molecule
is assembled from three reactants using two reaction steps. Details
related to the synthon dataset are given in Section
S\ref{sec:s-synthon-db}.

When generating each synthon structure, we adopt a graph-based approach
similar to our previous work\cite{Li.2021}. Specifically, we treat
the synthon structure as a 3D graph, and it is generated by iteratively
refining the graph structure. At each step, the model either adds a new
atom or a new bond or performs other operations such as backtracking. We
also introduce various improvements compared to the previous method,
including a more detailed treatment of ring generation. Specifically,
before generating each ring structure, DeepLigBuilder+ first specify the
size of the ring, as well as the location the ring will be closed. This
can better guide the generation process and can avoid potential issues
when the user changes the synthon dataset (see Figure
\ref{fig:s-ring-generation}). The generation scheme is detailed in
Section S\ref{sec:s-mol-gen}.

During generation, we need to constrain the synthon structure to the
space of purchasable building blocks. To achieve this, we perform
step-wise masking of the action space so that the generated structure
will not leave the space of purchasable synthons. The mask is
constructed by querying a prefix tree of synthon structures built from
the building block dataset, as detailed in Section
S\ref{sec:s-prefix-tree}. This new method of introducing chemical
constraints offers better scalability compared to previous
approaches\cite{bradshaw2020barking,Gao.2021}, which usually
requires a scan through the entire set of building blocks to generate
the next action.

DeepLigBuilder+ uses an SE(3) equivariant transformer to decide which
action to perform at each step of generation. Specifically, we convert a
3D molecular graph as a sequence of actions that are used to generate
its structure. This is equivalent to the concept of a ``sentence'' in
NLP-related tasks. Correspondingly, each action represents a ``word'' in
the sentence. The network adopts an encoder-decoder architecture, which
is used to translate the input information, including previously
generated synthons and pharmacophores (discussed below), into new
synthon structures, as shown in Figure \ref{fig:network}. To incorporate
3D information in an equivariant manner, we attach a 3D coordinate
system to the focused atom after each action and use invariant point
attention (IPA)\cite{Jumper.2021} to communicate information between
actions. We also use relative 3D positional encoding to express the
spatial relationship between action pairs. For network training, we
assemble a drug-like set of synthesizable molecules from the Enamine
building blocks, and use the 3D structures of these molecules generated
by RDKit to train the model. More information related to the network and
its training are given in Section S\ref{sec:s-network} and Section
S\ref{sec:s-dataset}.

\begin{figure}[t!]
\hypertarget{fig:mcts}{%
\centering
\includegraphics[width=1\textwidth,height=\textheight]{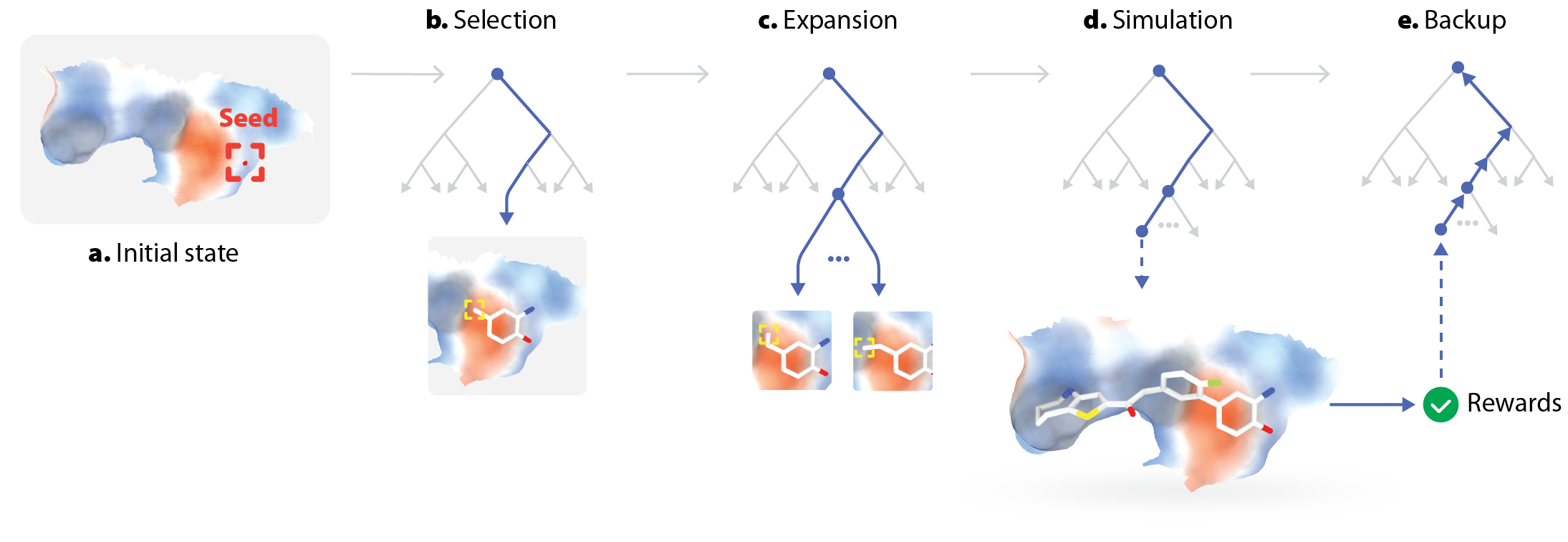}
\caption{DeepLigBuilder+ uses Monte Carlo tree search (MCTS) to achieve
structure-based molecule generation given the pocket structure. The
search starts from a user-provided seed atom (\textbf{a}). In this
figure, we use the ATP-binding pocket of BTK as an example. At each
step, the model first selects a promising node from the look-ahead tree
(\textbf{b}), then expands the tree by enumerating possible actions that
can be performed on the 3D molecule (\textbf{c}). Next, the model
selects an expanded state and uses the conditional transformer to
perform the rollout(\textbf{d}). Finally, the generated molecule is
evaluated using the Smina score, and the reward is backpropagated to
parent nodes to update the Q-value
estimates(\textbf{e}).}\label{fig:mcts}
}
\end{figure}

To introduce 3D information of targets, Monte Carlo tree search (MCTS),
a widely used algorithm in reinforcement learning, is applied to
optimize the molecule structure inside the pocket (Figure
\ref{fig:mcts}). We use a search method similar to
MENTS\cite{xiao2019maximum}, with custom modifications described in
Section S\ref{sec:s-mcts}, and a reward function based on the Smina
score\cite{Koes.2013}. The generated 3D pose by DeepLigBuilder+ is
directly used for scoring, eliminating the time-consuming docking step.

To enhance the performance of MCTS, we developed a pharmacophore and
shape-conditioned transformer as its rollout policy. Pharmacophore
models represent abstracted interaction patterns that can be used to
explain the bioactivity of ligands and are widely used in computer-aided
drug design\cite{Giordano.2022}. Shape information can help the
model by constraining the molecule to match the geometry of the pocket.
Those information are coded using SE(3) equivariant representations
discussed in Section S\ref{sec:s-network}. For model training, a dataset
of pharmacophore-ligand pairs is created by aligning synthesizable 3D
molecules to pharmacophore features extracted from PDBBind ligands
(Section S\ref{sec:s-dataset}). The conditional rollout policy offers
significant speed ups for MCTS search, as demonstrated in the following
sections.

\begin{figure}
\hypertarget{fig:unconditional}{%
\centering
\includegraphics[width=0.8\textwidth,height=\textheight]{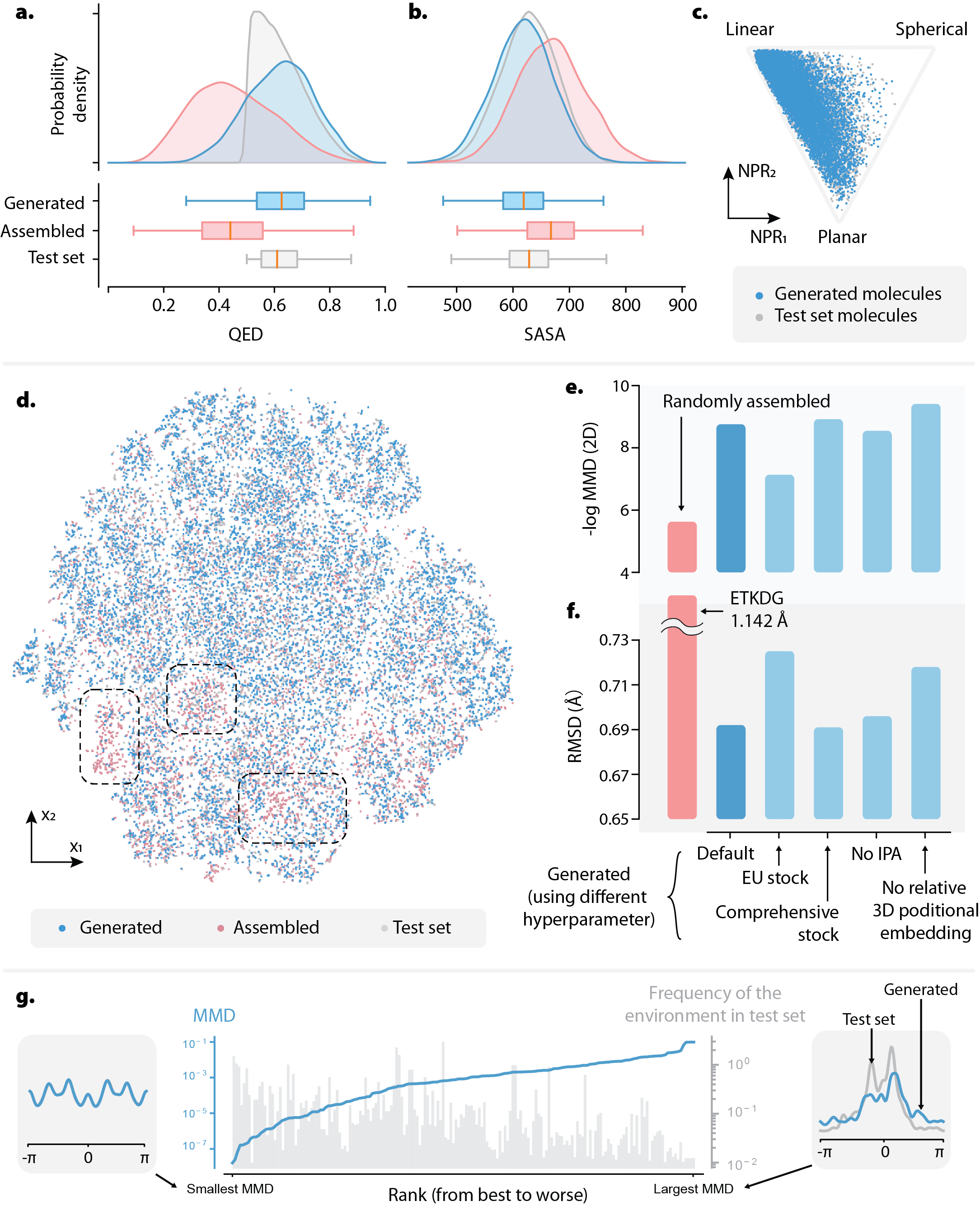}
\caption{The performance of the unconditional generative network.
\textbf{a-b.} The distribution of QED (\textbf{a.}) and SASA
(\textbf{b.}) among generated (blue), randomly assembled (red), and test
set (grey) molecules. \textbf{c.} The distribution of molecular shape.
\textbf{d.} A visualization of the distribution of 2D Morgan
fingerprints among generated (blue), randomly assembled (red), and test
set (grey) molecules. Boxes with dashed borders show locations in which
randomly assembled (non-druglike) molecules are enriched. \textbf{e.}
The 2D MMD values (negative log scale) of generated molecules using
different model configurations. The red bar shows the values calculated
using randomly assembled molecules. \textbf{f.} The RMSD values of
generated molecules using different model configurations. The red bar
shows the RMSD values for conformers generated using
ETKDG.}\label{fig:unconditional}
}
\end{figure}

\hypertarget{sec:results}{%
\section{Results}\label{sec:results}}

\subsection{Performance of the unconditional generative model}

We first evaluate the performance of the transformer network in the
unconditional setting, in a manner similar to our previous
work\cite{Li.2021}. Specifically, we investigate whether the model
is capable of generating drug-like and synthesizable molecules with
valid 3D structures. Several generated molecules by the network are
shown in Figure \ref{fig:s-uncond-samples}. A visual inspection of these
molecules reveal that they all adopt reasonable 3D conformations. Local
geometries are correctly structured based on the hybridization state of
their atomic environments. Neighbors of sp2-atoms are planarized, while
that of sp3-atoms form tetrahedron structures. Also, the overall
conformations generated are relaxed and contain no significant clashes.
Those observations will be later confirmed using quantitative evaluation
metrics (see Section \ref{sec:3d-quality}).

DeepLigBuilder+ is unique in that synthetic paths are also generated for
each molecule along with its 3D structures. Figure
\ref{fig:network}\textbf{e} shows the proposed route for synthesizing
structure \textbf{1}, which contains explicit purchasable reactants with
Enamine IDs. In this way, the synthesis routes of generated molecules
can be greatly simplified, potentially reducing the complexity of
wet-lab evaluations of those molecules.

\subsubsection{Distribution of 2D and 3D molecular properties}

To verify whether the model is capable of generating molecules with
desirable drug-like properties, 10,000 structures are sampled from the
model and several important 2D properties are calculated for each
molecule. The properties include molecular weight, LogP, the number of
rotatable bonds (ROT), hydrogen bond donors (HBD) and acceptors (HBA),
as well as QED\cite{Bickerton.2012}, which is a widely used metric
for drug-likeness estimation. The distributions of those properties are
visualized in Figure \ref{fig:s-prop} and Figure
\ref{fig:unconditional}. It can be seen that the distribution of most
properties matches well between the generated (blue) and test set
molecules (grey). A majority of generated molecules (85\%) have a QED
value larger than 0.5, as shown in Figure
\ref{fig:unconditional}\textbf{a}, indicating high drug-likeness.

We also compared the result with molecules randomly assembled from
reactants without any drug-likeness filters (shown in red). In general,
the property distribution of these randomly assembled molecules differs
significantly from that of the drug-like test set, with high molecular
weight and low drug-likeness. To offer a more quantitative evaluation,
we calculate the sum of squared differences between the mean and the
standard deviation statistics between generated and test-set molecules,
as shown in Table \ref{tbl:s-prop-2d}. Mathematically, this metric is
equivalent to the Wasserstein distance between Gaussian approximations
of two distributions. The property distributions of the generated
molecules are indeed more similar to the drug-like test set, compared to
the randomly assembled molecules, confirming that the model can indeed
significantly enrich drug-like molecules.

Similarly, several 3D molecular descriptors are calculated to examine
the method's ability to model 3D properties. Figure
\ref{fig:unconditional}\textbf{b} and Figure
\ref{fig:s-prop}\textbf{g-i} shows the distribution of
solvent-accessible surface areas (SASA\cite{Mitternacht.2016}),
Polar SASA and the radius of gyration(\(R_g\)\cite{Todeschini.2003})
for the generated and test set molecules. Like the 2D case, a close
match between the property distributions is found. In addition, we
visualized the shape distribution of these molecules using normalized
PMI ratios (NPRs\cite{Sauer.2003}), as shown in Figure
\ref{fig:unconditional}\textbf{c}. It can be seen that the generated
molecules are enriched in the linear region, while tilted towards the
planar region, following the distribution of test set molecules.
Randomly assembled molecules show a very different distribution in most
3D properties, and quantitative measurements shown in Table
\ref{tbl:s-prop-3d} confirm that molecules generated by the model share
higher similarity in 3D properties with the validation and test set
compared with assembled ones.

\subsubsection{The ability for the model to correctly model the
drug-like chemical space}

To access the network's ability to model the drug-like space of
synthesizable molecules, we visualize the distribution of 2D and 3D
structures for the generated, test set, and assembled molecules, as
shown in Figure \ref{fig:unconditional}\textbf{d} and Figure
\ref{fig:s-tsne}. Morgan and USRCAT\cite{Schreyer.2012} fingerprints
are used to represent 2D and 3D molecule structures and
t-SNE\cite{van2008visualizing} is used for dimension reduction. The
figures suggest that the overall distribution matches well between the
generated (blue) and test set (grey) molecules. On the other hand, there
are regions enriched with randomly assembled molecules (shown as dashed
boxes in Figure \ref{fig:unconditional}\textbf{d}), which likely
represent locations in chemical space featuring low drug-likeness.

To offer a more quantitative evaluation, we use maximum mean discrepancy
(MMD\cite{Gretton.2012}) to measure the overlap in the chemical
space for generated and test-set molecules, as done in previous
works\cite{Li.2019,Li.2021}. MMD is a metric used to determine the
dissimilarity between two probability distributions. Here, we also use
Morgan and USRCAT fingerprints as representations for MMD calculation in
2D and 3D chemical space. Results are detailed in Table \ref{tbl:s-mmd}
and Figure \ref{fig:unconditional}\textbf{e}. We can see the MMD value
is lower for molecules generated by the model compared with those
randomly assembled from the building blocks, indicating higher
similarity to the test set. We can conclude that after training, the
network can enrich the output to the drug-like portion of the chemical
space.

\hypertarget{sec:3d-quality}{%
\subsubsection{The quality of generated 3D
structures}\label{sec:3d-quality}}

An important aspect of the model's performance is its ability to
generate valid 3D structures of molecules. For evaluation, we first
measure its ability to correctly model the distribution of torsion
angles in different environments. The environments are described using
torsion SMARTS patterns by Schärfer et al.\cite{Scharfer.2013} We
compare the torsion distribution between the generated and test-set
molecules for each environment using MMD, and rank the value from lowest
to highest, as shown in Figure \ref{fig:unconditional}\textbf{g}. Since
the exact value of MMD lacks interpretability, we visually inspect the
torsion distribution for the environment with the highest and lowest MMD
values. Results show a good match between the distribution even for the
case with the highest MMD value, indicating that the model can correctly
construct the local geometries of molecules.

Next, we evaluate the global conformation quality using RMSD calculated
after relaxing the generated molecules with the MMFF94s forcefield. The
average RMSD values are reported in Table \ref{tbl:s-mmd} and Figure
\ref{fig:unconditional}\textbf{f}, while the distribution is shown in
Figure \ref{fig:s-rmsd}. Conformations produced by the model generally
have low RMSD values, with an average of 0.69\(\text{\AA}\). As a reference,
the RMSD value after relaxation for conformations generated by ETKDG
using the same set of topological structures is 1.14\(\text{\AA}\), a
significantly higher value. ETKDG is a conformation generation method
based on distance geometry and the empirical distribution of torsion
angles. This method is initially developed to provide a faster
alternative to MMFF94s optimization.

\begin{figure}
\hypertarget{fig:btk}{%
\centering
\includegraphics[width=0.9\textwidth,height=\textheight]{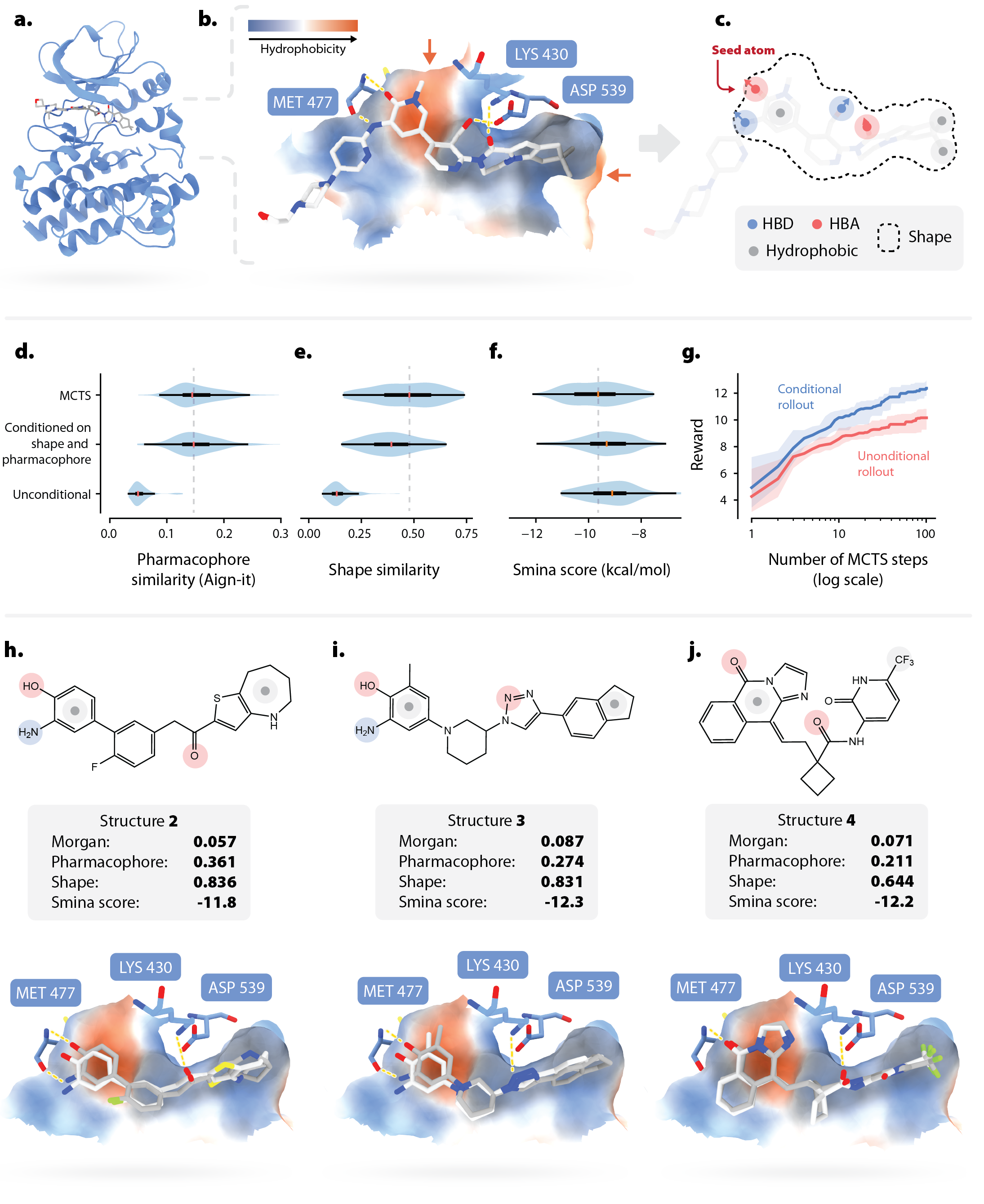}
\caption{\textbf{a.} The structure of BTK kinase domain in complex with
GDC-0853 (Fenebrutinib). \textbf{b.} The interaction between GDC-0853
and the ATP-binding pocket of BTK. \textbf{c.} The pharmacophore and
shape condition extracted based on the interaction pattern. As well as
the seed atom used for molecule growth. \textbf{d-f.} The distribution
of: \textbf{d.} the similarity with the given pharmacophore, \textbf{e.}
the similarity with the given shape, \textbf{f.} Smina docking scores
inside the ATP-binding pocket of BTK, among generated molecules.
\textbf{g.} The best rewards among all generated molecules at each step
of MCTS. \textbf{h-j.} Several generated molecules with high predicted
binding affinity. The conformation generated by the model is shown in
grey, and that produced by redocking the molecule is shown in
white.}\label{fig:btk}
}
\end{figure}

\subsection{Case study: designing inhibitors targeting BTK's ATP-binding
pocket}

As mentioned previously, DeepLigBuilder+ can perform pocket-based 3D
molecule design by combining a pharmacophore and shape-conditioned
transformer network with MCTS. To demonstrate its performance, we use
DeepLigBuilder+ to design inhibitors that bind to the ATP-binding pocket
of BTK. Bruton's tyrosine kinase (BTK) plays important roles in the
signal transmission in B-cells\cite{Satterthwaite.1998} and is
related to a series of related diseases, including B-cell
malignancies\cite{Singh.2018} and autoimmune
diseases\cite{Zhang.2021x}. Due to its importance, a variety of
compounds have been developed to inhibit BTK, mostly targeting its
ATP-binding pocket, with several of them approved for clinical
use\cite{Tasso.2021}. However, all currently approved inhibitors
bind to BTK covalently via Cys481, which may cause off-target effects by
binding with other kinases with Cys481-like
residues\cite{Sibaud.2020}. Resistant mutations on Cys481 can also
reduce their clinical effect\cite{Woyach.2014}. Non-covalent binders
can in theory avoid those disadvantages, and in this section, we attempt
to apply DeepLigBuilder+ for the design of non-covalent BTK inhibitors.

Several non-covalent BTK inhibitors are currently under clinical
investigation. Here, we extract pharmacophore features and attempt to
use DeepLigBuilder+ to generate new molecules with novel structures.
Figure \ref{fig:btk}\textbf{a} shows the structure of the kinase domain
of BTK (PDB ID: 5vfi) complexed with GDC-0853 (Fenebrutinib), a potent
BTK inhibitor (Ki=0.91nM) currently under clinical
trial\cite{Crawford.2018}. GDC-0853 utilizes several important
interactions inside the ATP-binding site such as hydrogen bonding with
Met477, Lys430, and Asp539, as well as the hydrophobic interactions in
the hinge region and the selectivity pocket (H3), as shown in Figure
\ref{fig:btk}\textbf{b}. Based on the information, we construct the
shape and pharmacophore information shown in Figure
\ref{fig:btk}\textbf{c}, and use it as a condition for DeepLigBuilder+.
Note that GDC-0853 also occupies a solvent-exposed region as seen in the
left part of Figure \ref{fig:btk}\textbf{b}. This region is not included
as input because we want to constrain the output molecule to a smaller
size so that it can be more ``lead-like'' and easier to be further
optimized. The oxygen atom of the carbonyl group is used as seed for
molecule growth due to its interaction with Met477.

After the shape and pharmacophore-based inputs are determined, we
combine the conditional generative model and MCTS to perform
structure-based molecule design, with details shown in Section
S\ref{sec:s-mcts}. First, we investigate whether the model can enrich
molecules based on the given condition. Figure \ref{fig:btk}\textbf{d-e}
shows that the conditional model can generate molecules with a better
match in pharmacophore and shape compared with unconditional methods.
Next, we evaluate the benefit of using the conditional rollout inside
MCTS. Figure \ref{fig:btk}\textbf{g} shows the best reward among all
generated molecules at each step of MCTS. It is shown that the
conditional rollout can help to speed-up MCTS search. We also evaluate
the benefit of using MCTS compared with direct sampling from the
conditional model. Figure \ref{fig:btk}\textbf{f} shows that MCTS can
help to improve the docking score of generated molecules, with a 0.40
kcal/mol improvement in mean values (comparing the first row against the
second row in Figure \ref{fig:btk}\textbf{f}). It also offers more
enrichment in the range of high binding affinity. 13\% of molecules
generated using MCTS have a smina score \textless{} -10 kcal/mol. The
value is reduced to 5\% if the molecules are sampled directly from the
conditional network.

To better demonstrate the molecule optimization process, we visualize
the search tree used in MCTS in Figure \ref{fig:s-mcts}. Due to space
constraints, the tree only contains states of the first synthon, and
nodes with a visit count less than 25 are dismissed. It is shown that
the model prioritizes the visits to states with higher Q-values, as
those states are expanded more often. States with lower Q-values (the
nodes on the left side of the tree) are less favored, due to reasons
such as bad 3D positions of the synthon anchors, as shown in Figure
\ref{fig:s-mcts}.

Several generated molecules with high predicted binding affinity are
shown in Figure \ref{fig:btk}\textbf{h-j}. A search in PubChem reveals
no highly similar molecules (Tanimoto similarity \textgreater{} 95\%),
indicating that those are indeed novel structures. Additionally, a
search in the ChEMBL database does not reveal any structurally related
molecules (Tanimoto similarity \textgreater{} 70\%), indicating that no
topologically similar molecules have been evaluated against BTK. The
topological similarity with the seed molecule extracted from GDC-0853
(Figure \ref{fig:btk}\textbf{c}) is also low. In contrast, in terms of
pharmacophore and shape, most input pharmacophore features are covered
inside these generated molecules, including hydrogen bonding with Lys430
and Met477 and hydrophobic interactions at the two ends of each
molecule. Synthesis paths are also proposed by DeepLigBuilder+ for each
generated molecule, making retrosynthetic analysis of generated
molecules easier. For example, Figure \ref{fig:btk-reaction}\textbf{a}
shows the proposed synthetic path for Structure \textbf{4}. The path for
Structure \textbf{2} and \textbf{3} are shown in Figure
\ref{fig:s-btk-reaction}. In summary, by combining MCTS with the
conditional generative model, DeepLigBuilder+ can enrich molecules with
high binding affinity based on the pharmacophore constraint.

\begin{figure}
\hypertarget{fig:phgdh}{%
\centering
\includegraphics[width=0.9\textwidth,height=\textheight]{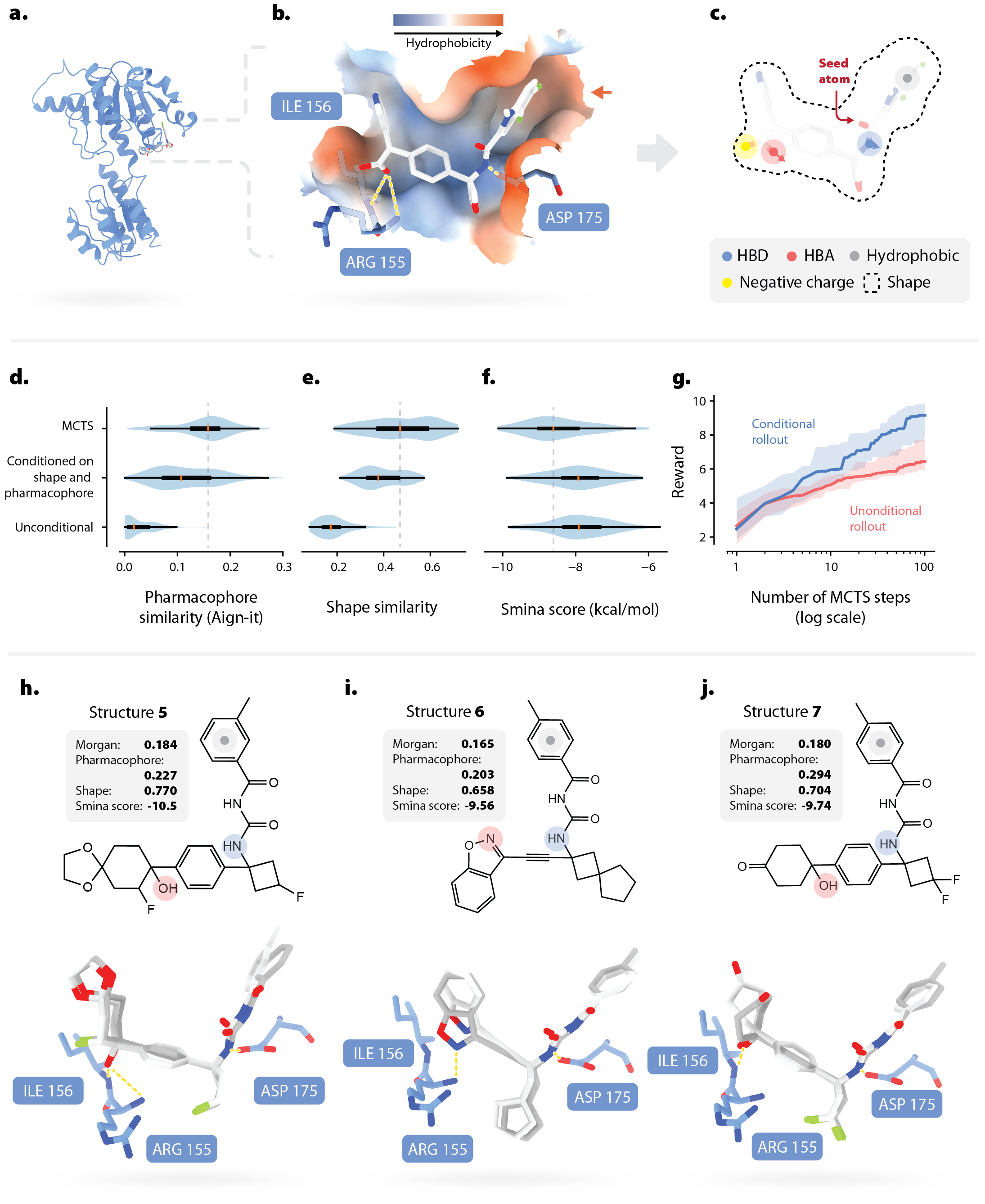}
\caption{\textbf{a.} The structure of PHGDH with compound \textbf{15}.
\textbf{b.} The interaction between compound \textbf{15} with the NAD+
binding pocket of PHGDH. \textbf{c.} The pharmacophore and shape
condition extracted based on the interaction pattern. As well as the
seed atom used for molecule growth. \textbf{d-f.} The distribution of:
\textbf{d.} the similarity with the given pharmacophore, \textbf{e.} the
similarity with the given shape, \textbf{f.} Smina docking scores inside
the NAD+ binding pocket of PHGDH, among generated molecules. \textbf{g.}
The best rewards among all generated molecules at each step of MCTS.
\textbf{h-j.} Several generated molecules with high predicted binding
affinity. The conformation generated by the model is shown in grey, and
that produced by redocking the molecule is shown in
white.}\label{fig:phgdh}
}
\end{figure}

\subsection{Case study: designing inhibitors targeting the NAD+ pocket
of PHGDH}

Targeting cancer metabolism represents an important strategy for cancer
drug development\cite{Faubert.2020}. Human phosphoglycerate
dehydrogenase (PHGDH), a key enzyme in the serine biosynthesis pathway,
has been demonstrated to have crucial roles in
tumorigenesis\cite{Zhao.2021}, making it a promising cancer-related
target. One strategy for targeting PHGDH is to design inhibitors that
bind to its NAD\(^+\) pocket. Multiple such inhibitors have been
reported in previous works, with most of them containing an indole-based
scaffold. In this case study, we use DeepLigBuilder+ to design potential
binders for the NAD+ pocket with novel structures.

Figure \ref{fig:phgdh}\textbf{a} shows the structure of PHGDH (PDB ID:
6plg) together with compound \textbf{15}, a potent inhibitor of the
target developed by Mullarky et al.\cite{Mullarky.2019}. Figure
\ref{fig:phgdh}\textbf{b} demonstrates the interaction between the
ligand and the target. The nitrogen atom in the amide group in compound
\textbf{15} acts as a hydrogen bond donor and interacts with Asp175. The
carboxyl group in compound \textbf{15} can form hydrogen bonds with
backbone nitrogen atoms. It also forms charge-charge interaction with
Arg155. The indole structure of compound \textbf{15} resides inside a
hydrophobic region, as shown by the orange arrow in Figure
\ref{fig:phgdh}\textbf{b}. A pharmacophore model is constructed based on
those interactions, as shown in Figure \ref{fig:phgdh}\textbf{c}. We use
the amide structure as the seed for molecule growth, as shown in Figure
\ref{fig:phgdh}\textbf{c} and Figure \ref{fig:s-compound-15}\textbf{b}.
The shape of compound \textbf{15} is also used as an input feature.

Next, we use DeepLigBuilder+ to generate molecules based on the
pharmacophore and shape information and the target structure. Similar to
the previous section, we first evaluate whether the conditional model
offers more enriched results based on the provided information. Indeed,
Figure \ref{fig:phgdh}\textbf{d-e} shows that molecules sampled from the
conditional transformer match better to the input pharmacophore(Figure
\ref{fig:phgdh}\textbf{d}) and shape(Figure \ref{fig:phgdh}\textbf{e})
compared with the unconditional model. Figure \ref{fig:phgdh}\textbf{g}
shows that introduction conditions help to accelerate the MCTS search,
as demonstrated by the blue curve. When evaluating the benefit of the
MCTS module, we found that MCTS search helps the model to generate
molecules with better pharmacophore and shape matches (Figure
\ref{fig:phgdh}\textbf{d-e}), and also helps to improve the docking
score of the result, with an average improvement of Smina score of 0.53
kcal/mol. 11\% of molecules generated using MCTS have a Smina score
\textless{} -9.5 kcal/mol, compared to the value of 2\% for those
directly sampled from the conditional model.

\begin{figure}
\hypertarget{fig:btk-reaction}{%
\centering
\includegraphics[width=1\textwidth,height=\textheight]{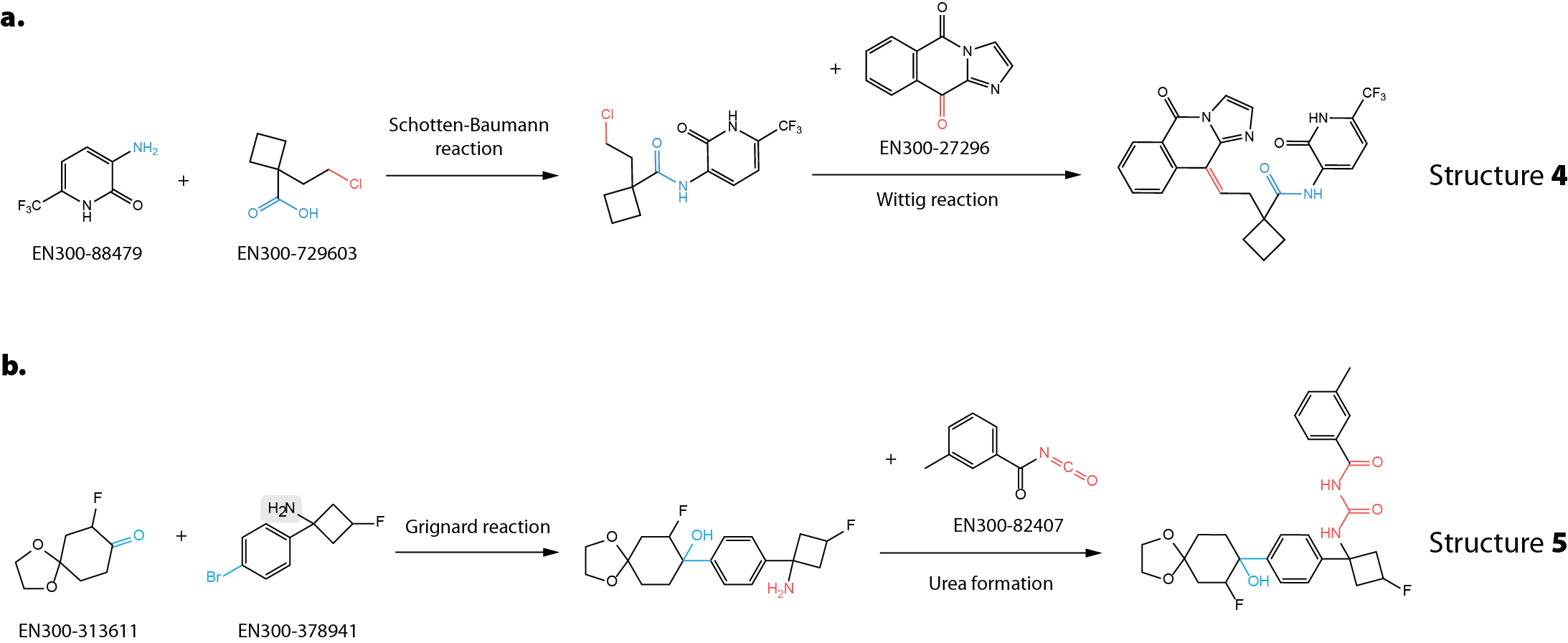}
\caption{\textbf{a.} The synthetic path of Structure \textbf{4} proposed
by DeepLigBuilder+. It involves a two-step process, which first connects
the amide bond using the Schotten-Baumann reaction, and then forms the
double bond using the Wittig reaction. Enamine IDs of reactants are also
given. Note that the Schotten-Baumann reaction requires an additional
activation step that transforms the carboxyl group into the acyl
chloride. Also, the Wittig reaction requires the formation of the ylide.
\textbf{b.} The synthetic path of Structure \textbf{5} proposed by
DeepLigBuilder+. The first two reactants are connected using the
Grignard reaction. The third reactant is connected by forming a urea
structure using the amine group and the isocyanate group. Note that
before the first step, reactant 2 needs to be transformed into the
Grignard reagent. Additionally, the amine group in reactant 2 (marked
grey) needs to be protected before carrying out other
reactions.}\label{fig:btk-reaction}
}
\end{figure}

Figure \ref{fig:phgdh}\textbf{h-j} shows several molecules generated by
DeepLigBuilder+ with high predicted binding affinity. The reaction paths
generated by the model are shown in Figure
\ref{fig:btk-reaction}\textbf{b} and Figure \ref{fig:s-phgdh-reaction}.
Those molecules have low topological similarities with compound
\textbf{15}, but share pharmacophore features such as hydrogen bond
donors that interact with Asp175 and acceptors that interact with Ile156
or Arg155. A search in PubChem does not reveal results with high
topological similarity with those molecules (Tanimoto similarity
\textgreater95\%), indicating that those are indeed novel structures.
Also, the ChEMBL dataset does not contain topologically related compound
records (Tanimoto similarity \textgreater70\%), which means that similar
molecules are not yet evaluated against PHGDH. Interestingly, those
molecules contain cyclobutane structures that are similar to the oxetane
structure in compound \textbf{15}. This structural motif helps to form a
turn in the molecule shape for better accommodation with the pocket, and
also creates a hydrophobic interaction with Ile177.

\subsection{Ablation studies and the effects of different
hyperparameters}

It is important to understand how different architectural and
hyperparameter choices affect the performance of the proposed model. In
this section, we demonstrate the impact of several important network
features and hyperparameters. Details about the configurations explored
are shown in Table \ref{tbl:s-hyperparam}. The performances of the model
under different configurations are shown in Table \ref{tbl:s-mmd}.

\subsubsection{The effect of changing the set of accessible building
blocks}

A major feature of DeepLigBuilder+ is its capability to suggest
synthetic paths with accessible building blocks along with its generated
molecules. However, the accessibility of building blocks is a constantly
changing factor. On one hand, due to technical advances, the number of
synthesizable building blocks is rapidly growing over the years. On the
other hand, in-stock supply of such building blocks may vary between
times and locations, and most require on-demand synthesis, increasing
the cost. Ideally, DeepLigBuilder+ should allow the user to choose a
building block set that fits their need, without the need to perform
re-training. Here, we simulate such scenarios and report how the choice
of building blocks impacts the quality of generated molecules.

To simulate the lack of in-stock availability for certain building
blocks, we restrict the building blocks to the EU stock, which is a
smaller set with 81,235 compounds. In terms of the quality of
topological structures, although Figure
\ref{fig:unconditional}\textbf{e} shows that there is an increase in the
MMD value after the restriction, Table \ref{tbl:s-prop-2d} confirms that
the generated molecules still maintain a high drug-likeness, with an
average QED value of 0.60, close to the value before changing the
building blocks. In terms of the quality of 3D conformation, Table
\ref{tbl:s-mmd} and Figure \ref{fig:unconditional}\textbf{f} indicates
an increase of RMSD from 0.69\(\text{\AA}\) to 0.72\(\text{\AA}\), but still much
lower than 1\(\text{\AA}\). From the results, we believe that although using a
smaller building block may indeed impact the performance of the network,
such an effect should be minor and still allows for regular application
of DeepLigBuilder+ in drug design tasks.

Then, to demonstrate how increasing the building block set may affect
the generation result, we expand to include the comprehensive catalog,
which contains more than 1 million (1,162,033) compounds, some may be
synthesized on-demand. Figure \ref{fig:unconditional}\textbf{e-f} shows
that this change induces little impact on the performance of the output.
Only 28.1\% of molecules generated have used the newly added building
blocks. In conclusion, we believe that DeepLigBuilder+ still offers
promising performance when the building block set is changed, but a
re-training may be required if we want to fully utilize newly added
building blocks.

\subsubsection{The effect of different ways to encode 3D structural
information}

Besides using the relative 3D positional encoding module to incorporate
3D information, DeepLigBuilder+ also uses invariant point attention
(IPA), which offers a geometrically-aware way to pass information
between atoms. To understand the benefit of including IPA, we disable
the two modules consecutively and investigate the impact of those
changes. The results are shown in Figure
\ref{fig:unconditional}\textbf{e-f} and Table \ref{tbl:s-mmd}. We found
that removing IPA has little impact on the quality of 3D conformation,
as measured by RMSD, showing that IPA is not essential for maintaining
3D structure quality. However, the quality of 2D structure, as measured
by 2D MMD, is reduced. If we remove the 3D positional embedding and keep
IPA, we observe a significant improvement in 2D structural quality, but
the RMSD value increased to 0.718\(\text{\AA}\).

The results above demonstrated a trade-off between 2D and 3D structural
quality, and that the two ways of including 3D information, IPA and
relative positional encoding, have different emphases on the two
aspects. One way to understand the result is to view IPA as a more
regularized way of encoding geometric information. Our previous work has
demonstrated that the model can not reliably generate correct 2D
structures if it overly relies on accurate 3D information since it
reduces the model's ability to recover from errors during
generation\cite{Li.2021}. The highly structured way to communicate
3D information in IPA can act as a form of regularization on how the
model uses the 3D information. On the other hand, there is no limitation
on how 3D relative positional embedding will be processed by the
network. Therefore, although the model can still generate accurate 3D
structures without IPA, a lack of regularization will reduce the quality
of 2D structures. Using both modules acts as a compromise, with an
improved 3D conformation quality and a balanced 2D structure quality.

\section{Conclusion}

We have developed a new de novo drug design tool, DeepLigBuilder+, that
generates synthesis-driven 3D molecules for a given target.
DeepLigBuilder+ uses geometric transformer combined with an MCTS-based
reinforcement learning module to navigate the space of synthesizable 3D
molecules to identify potential bioactive compounds. This method aims to
address two major challenges faced by deep molecule generative models:
(1) the design of 3D molecules based on 3D constraints, and (2) the
design of molecules with high synthetic accessibility. DeepLigBuilder+
has shown promising performances in overcoming these challenges.
DeepLigBuilder+ is capable of generating 3D molecules with high
drug-likeness and geometric quality under the synthetic accessibility
constraint. In the case study related to BTK and PHGDH, DeepLigBuilder+
significantly enriches molecules with high docking scores and favorable
interaction patterns with the target pocket, using the 3D information
provided. For each generated molecule, DeepLigBuilder+ proposes a
synthetic route with explicit reactions and building blocks that can be
directly queried from the supplier, making retrosynthetic analysis much
easier.

DeepLigBuilder+ takes advantage of recent developments in 3D generative
networks\cite{Li.2021} and the idea of synthetically aware de novo
design\cite{Coley.2020}. To restrict the model to the chemical space
of synthesizable molecules, we develop a method that calculates a
stepwise constraint of the generation trajectory to ensure that it leads
to purchasable building blocks. Compared to other approaches that
require a fragment-based generation scheme, our method can in theory be
applied to various atom-based molecule generative models. In addition,
it offers better scalability to large building block datasets by
organizing them into a tree-based structure and avoids full database
scans at each step. To achieve structure-based generation, we
constructed a new dataset of pharmacophore-ligand pairs using
large-scale 3D alignment of molecules, and then use it to develop a
novel SE(3)-equivariant transformer conditioned on 3D information. This
network is then combined with MCTS as the rollout policy, and it is
demonstrated that the combination results in a significant improvement
in the search speed of MCTS.

DeepLigBuilder+ could be improved in the following aspects in the
future. First, the present molecule assembling process relies on simple
SMARTS rules, which may be limited in precision. We are planning to
include a more dedicated model for yield and selectivity prediction so
that we can further improve the synthesizability of the assembled
molecules by the model. Second, we are planning to update the reaction
set to include broader, more modern reactions. In addition, the current
version of DeepLigBuilder+ requires user-provided seed structures for
molecule growth, and we are planning to develop methods to automate the
seed selection process. Finally, we are planning to build a transformer
model that is directly conditioned on the 3D pocket structure, without
the need for pharmacophore extraction. In summary, due to its unique
capability of generating highly synthesizable molecules with 3D
structures, DeepLigBuilder+ provides a powerful tool to generate
bioactive molecules and to accelerate the process of structure-based
drug design.

\section{Acknowledgements}

This work has been supported in part by the National Natural Science
Foundation of China (22033001). We would like to thank Yuhao Ren and
Kangjie Lin for their kind advice on the synthetic accessbility of
generated molecules. We also appreciate Alibaba Cloud for providing the
EFLOPS computation platform for the GPU-based network training.

%% file: si.tex
\hypertarget{sec:s-method}{%
\section{Supplementary Methods}\label{sec:s-method}}

\hypertarget{sec:s-synthon-db}{%
\subsection{Constructing the synthon dataset}\label{sec:s-synthon-db}}

We use the building block sets provided by Enamine as the set of
purchasable reactants. The global stock, which contains 238,980
compounds at the time of access (July 2022), is used to assemble the
training set molecules. To investigate the impact of changing available
reactants on the model's performance, we also downloaded the EU stock
and comprehensive catalog, which contains 81,235 and 1,162,033 molecules
respectively (by Oct 2022).

For reactions, we use the 58 SMARTS rules collected by Hartenfeller et
al.\cite{Hartenfeller.2011}, which represents a set of robust
chemical reactions relevant to drug design. Based on the reaction set,
we constructed a set of SMARTS rules to convert reactants to synthons.
The conversion is performed using the following procedure:

\begin{enumerate}
\def\labelenumi{(\arabic{enumi})}
\item
  For a given building block and reaction rule, we predict the product
  structure using RDKit. If the reaction involves two reactants, the
  other reactant is set to be a minimum structure with the required
  function group.
\item
  The substructure in the product molecule that corresponds to the
  building block is extracted. The open valences resulting from the bond
  break in the extracted substructure are labeled with the reaction type
  and its role in the reaction.
\item
  If the substructure extraction results in multiple valences, this
  indicates that new rings are formed after the reaction. In this case,
  we either include the new ring structure inside this synthon or leave
  the ring to the synthon corresponding to the other reactant.
\end{enumerate}

\begin{figure}
\hypertarget{fig:s-synthon-rules}{%
\centering
\includegraphics[width=1\textwidth,height=\textheight]{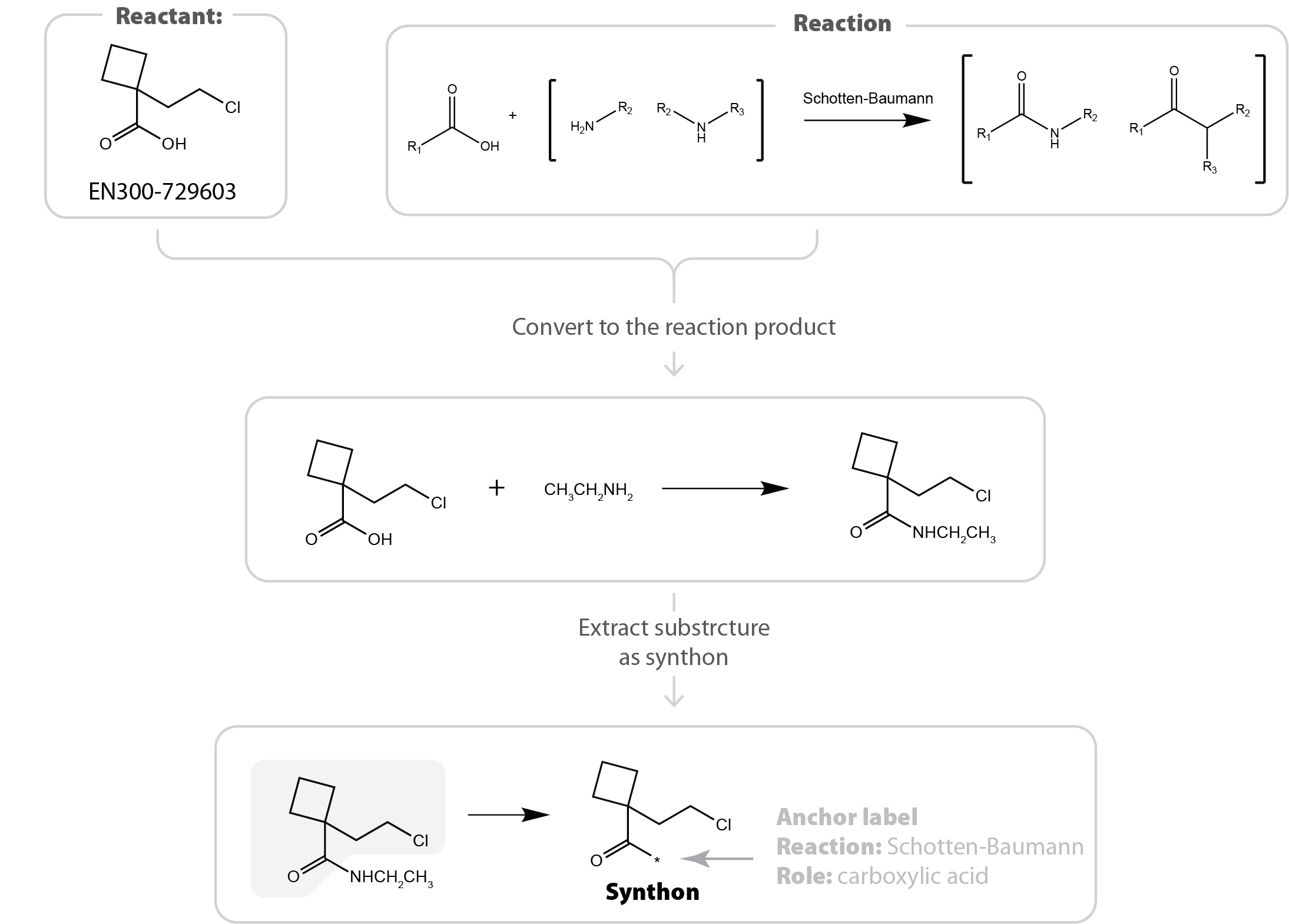}
\caption{The process of converting a reactant in the building block
dataset into a synthon.}\label{fig:s-synthon-rules}
}
\end{figure}

\begin{figure}
\hypertarget{fig:s-synthon-rules-ring}{%
\centering
\includegraphics[width=1\textwidth,height=\textheight]{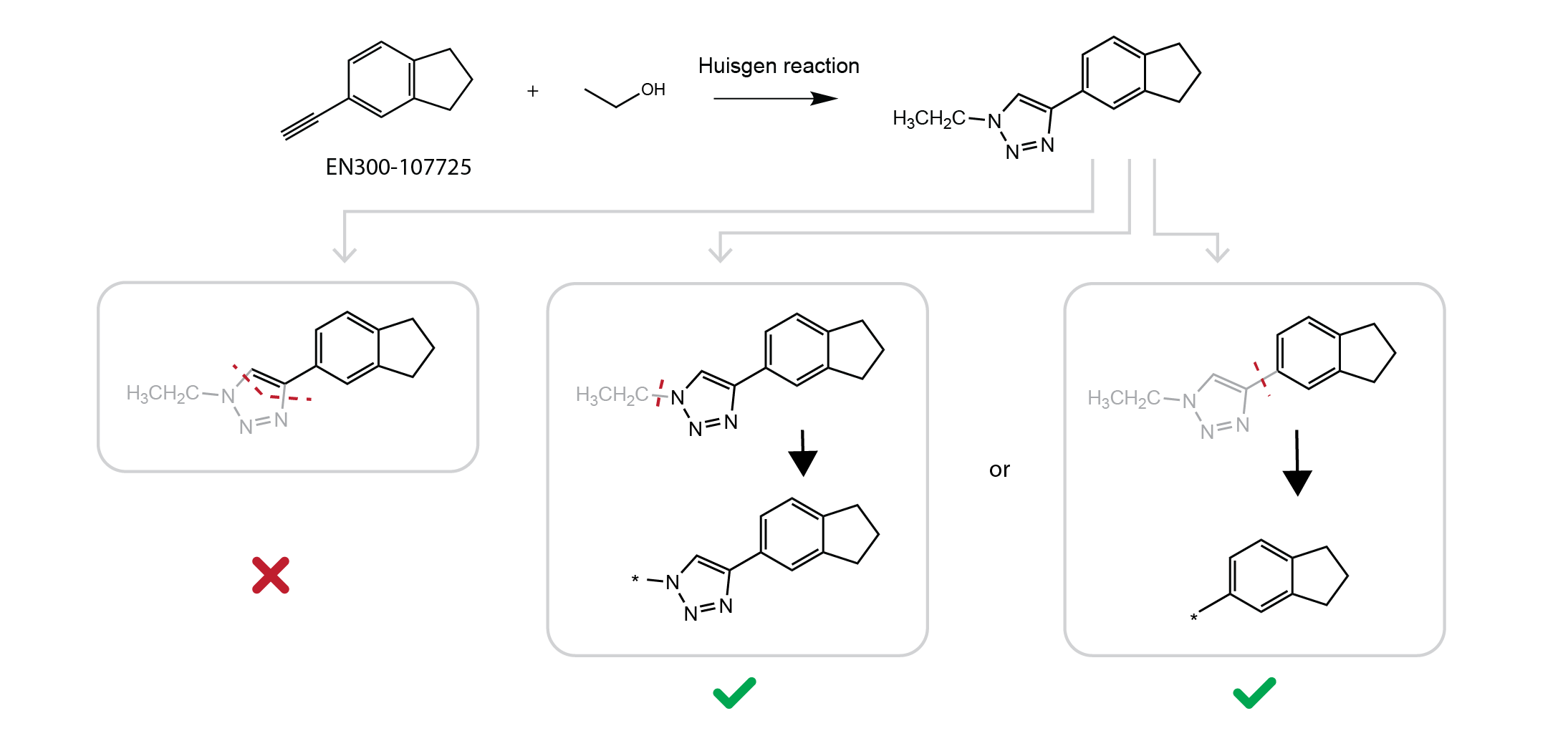}
\caption{Special treatment is needed when converting reactants to
synthons when the reaction involves ring
formation.}\label{fig:s-synthon-rules-ring}
}
\end{figure}

A visual demonstration of this process is given in Figure
\ref{fig:s-synthon-rules} and Figure \ref{fig:s-synthon-rules-ring}. To
simplify the generative model, we require that each reaction will result
in at most one open valence in synthon structures, and such valence must
correspond to a single bond in the product molecule. Most reactions
satisfy this requirement. As a result, we kept 52 reactions and
constructed 108 SMARTS rules for converting reactants to synthons.

The Enamine building blocks are then converted to synthons using those
rules. For each building block, we enumerate all possible synthon
structures by iterating through the reaction rules. At most two
reactions are allowed to happen in one reactant. To ensure that the
generated synthons are fragment-like and relevent to drug discovery, we
apply the following rules to filter the results:

\begin{enumerate}
\def\labelenumi{(\arabic{enumi})}
\item
  Element types of atoms inside each molecule are restricted to the set
  \{C, O, N, P, S, F, Cl, Br, I\} and bond types are restricted to
  single, double, triple, and aromatic bonds.
\item
  Synthons are required to be ``fragment-like'' based on the rule of two
  (Ro2)\cite{Goldberg.2015}.
\item
  Each fragment can have at most 4 rings, and the size of each ring
  should not be larger than 7.
\item
  Since we are focusing on designing non-covalent binders, we filter
  structures with the potential of forming covalent bonds with the
  protein, using a set of SMARTS rules \cite{London.2014}.
\end{enumerate}

The filtering results in a dataset containing 241,310 synthons from the
global stock, 103,385 synthons from the EU stock, and 783,195 synthons
from the comprehensive catalog. The synthons from the global stock are
later used to assemble the training set molecules (See Section
S\ref{sec:s-dataset}).

\hypertarget{sec:s-mol-gen}{%
\subsection{Performing molecule generation}\label{sec:s-mol-gen}}

DeepLigBuilder+ generates 3D molecules using a synthon-based method.
Each molecule is composed of 3 synthon structures, which is equivalent
to combining three building blocks with two reaction steps. When
generating each synthon, DeepLigBuilder+ uses a graph-based approach
similar to our previous work\cite{Li.2021}. Specifically, the model
generates 3D synthon structures by producing molecular graphs. We write
the output graph as \(G=(V, E, A, B, X)\), where \(V\) and \(E\) are the
set of nodes (atoms) and edges (bonds), \(A=\{a_v\}_{v\in V}\) and
\(B = \{b_{uv}\}_{\{u, v\}\in E}\) are labels representing the type of
each atom and bond, and \(X=\{\mathbf{x}_v\}_{v\in V}\) are the 3D
coordinates of each atom.

When generating each synthon, the model starts with an empty graph
\(G_0=(\empty, ..., \empty)\), iteratively updates its structure
\(G_t=a_t(G_{t-1})\), and outputs the graph as a new synthon fragment
when it is ready. During generating, additional information is attached
to each node in the graph to record the generation history, which
includes:

\begin{enumerate}
\def\labelenumi{(\arabic{enumi})}
\item
  The currently focused node, denoted as \(v^*_t\), where \(t\)
  represents the step ID. The definition of a ``focused'' node is
  similar to that in our previous works \cite{Li.2021}. Briefly, all
  edits to the molecular graph, whether to add new atoms or new bonds,
  happen on the focused node.
\item
  The parent of each node \(P_t=\{p_v\}_{v\in V_t}\). A node \(p_v\) is
  called the ``parent'' of another node \(v\) if \(p_v\) is the focused
  node when \(v\) is generated. Note that the parent-child relationship
  induces a spanning tree of the molecular graph \(G_t\), which can be
  used to calculate tree-based distances between atoms (or nodes) in the
  graph as input features to the transformer network (as detailed in
  Section S\ref{sec:s-network}).
\end{enumerate}

We denote the state of the molecular graph at step \(t\) with the
additional information as
\(G_t'=(V_t, E_t, A_t, B_t, X_t, v_t^*, P_t)\). The graph structure is
iteratively modified based on actions \(a_t\) sampled from the neural
network. The following types of actions are allowed during generation:

\begin{enumerate}
\def\labelenumi{(\arabic{enumi})}
\item
  Initialization, which adds the first atom to the molecular graph;
\item
  Append, which attaches a new atom to the focused atom using a new
  bond. For this action, the model needs to decide the type of the new
  atom and bond, as well as the 3D position of the new atom. When
  generating the position, the model uses a spherical coordinate frame
  attached to the focused atom, following our previous work
  \cite{Li.2021}. Besides the element type of the new atom, the
  model also needs to decide whether there will be branches on this atom
  and whether the atom will be a target of future ring closure, similar
  to Ahn et al.\cite{ahn2021spanning}
\item
  Backtracking, which requires the model to move the focused atom to its
  closest ancestor that allows branching. The generation terminates if
  the focused node has no parents.
\item
  Search loop target. This action indicates that a new ring will be
  formed, and the model should examine the closest ancestor of the
  focused node to see whether it is a suitable target during the ring
  closure. The network can generate this action multiple times until a
  suitable target is found for ring closure.
\item
  Start loop. This action happens after a ``search loop target'' action
  when the appropriate target of ring closure is found. In this action,
  the model determines the size of the ring. After this action, a series
  of ``append'' actions should be issued by the model to complete the
  ring formation.
\item
  Close loop. This action happens after all ring atoms have been
  generated, and the model is ready to connect the ring to the target
  atom determined in the ``search loop target'' step. The model decides
  the type of bond used to close the ring during this action.
\end{enumerate}

\begin{figure}
\hypertarget{fig:s-state-machine}{%
\centering
\includegraphics[width=1\textwidth,height=\textheight]{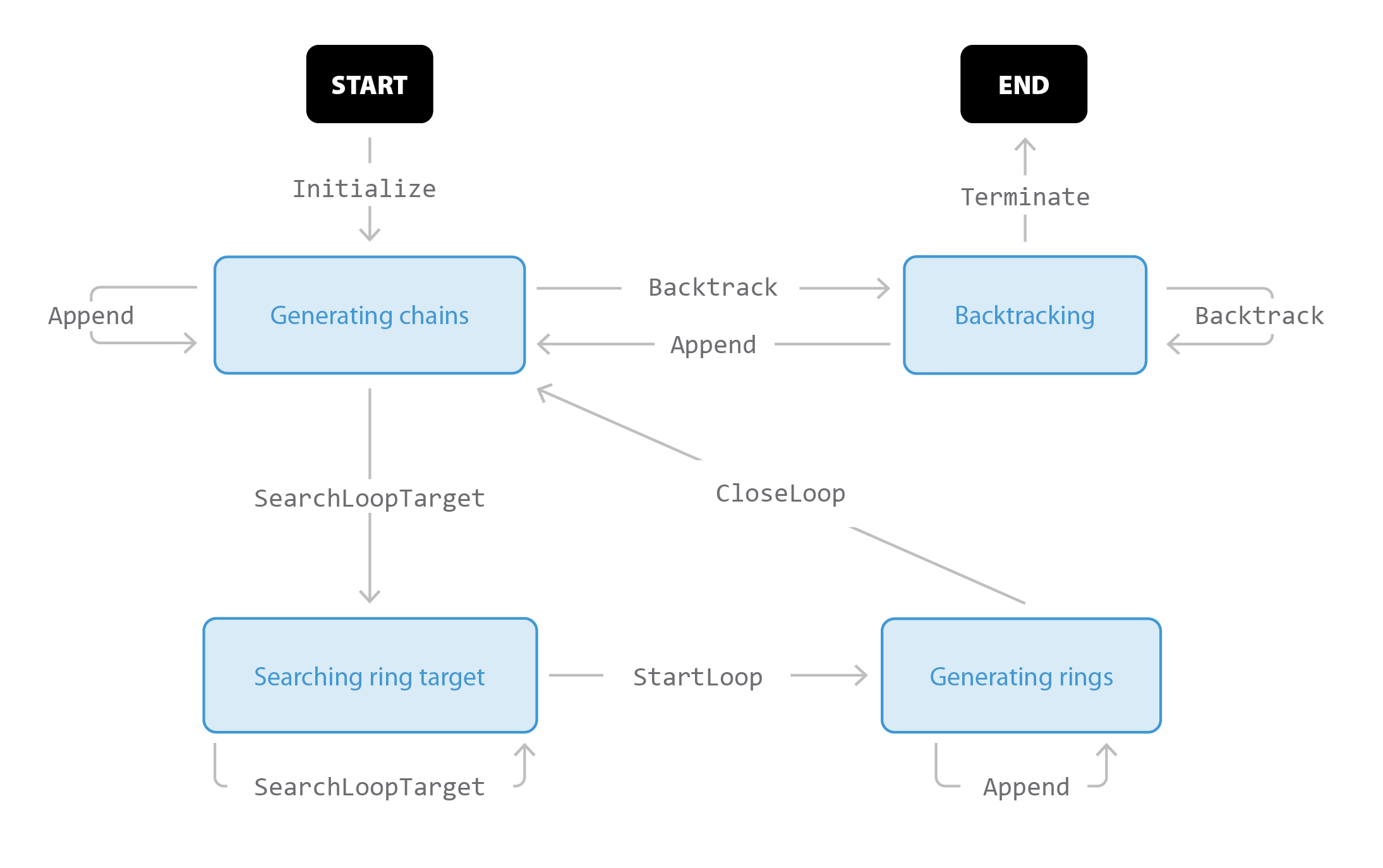}
\caption{A representation of the molecule generation process using a
finite state machine. From the figure, we can see that the model moves
back and forth between ring generation (represented as the ``searching
ring target'' state and the ``generating ring'' state) and chain
generation (represented as the ``generating chains'' state and the
``backtracking'' state).}\label{fig:s-state-machine}
}
\end{figure}

\begin{figure}
\hypertarget{fig:s-generation-process}{%
\centering
\includegraphics[width=1\textwidth,height=\textheight]{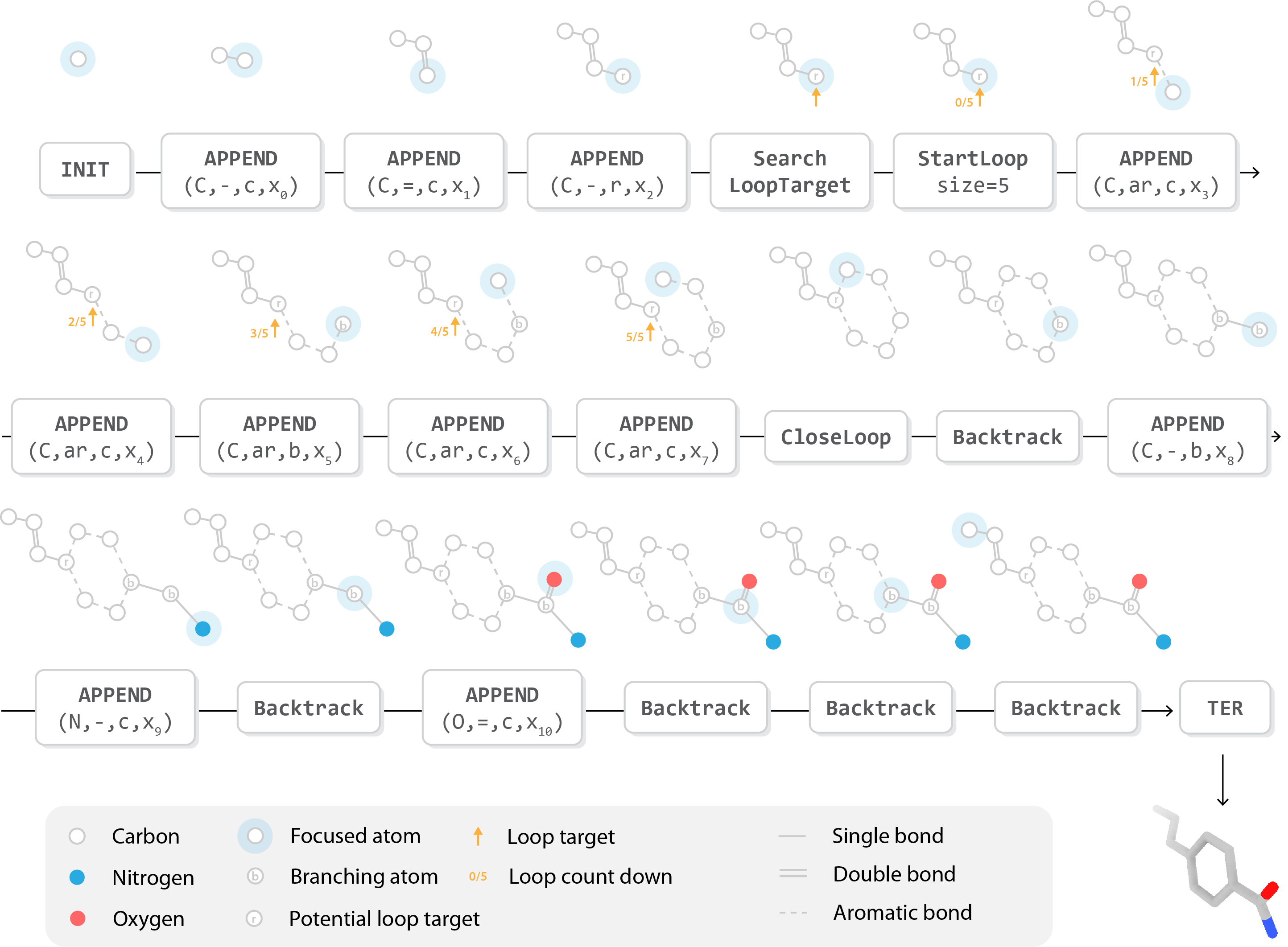}
\caption{The process of generating a 3D molecule using
DeepLigBuilder+}\label{fig:s-generation-process}
}
\end{figure}

The process of molecule generation can be represented using a finite
state machine, as shown in Figure \ref{fig:s-state-machine}. A full path
for generating an example molecule is shown in Figure
\ref{fig:s-generation-process}.

\begin{figure}
\hypertarget{fig:s-ring-generation}{%
\centering
\includegraphics[width=1\textwidth,height=\textheight]{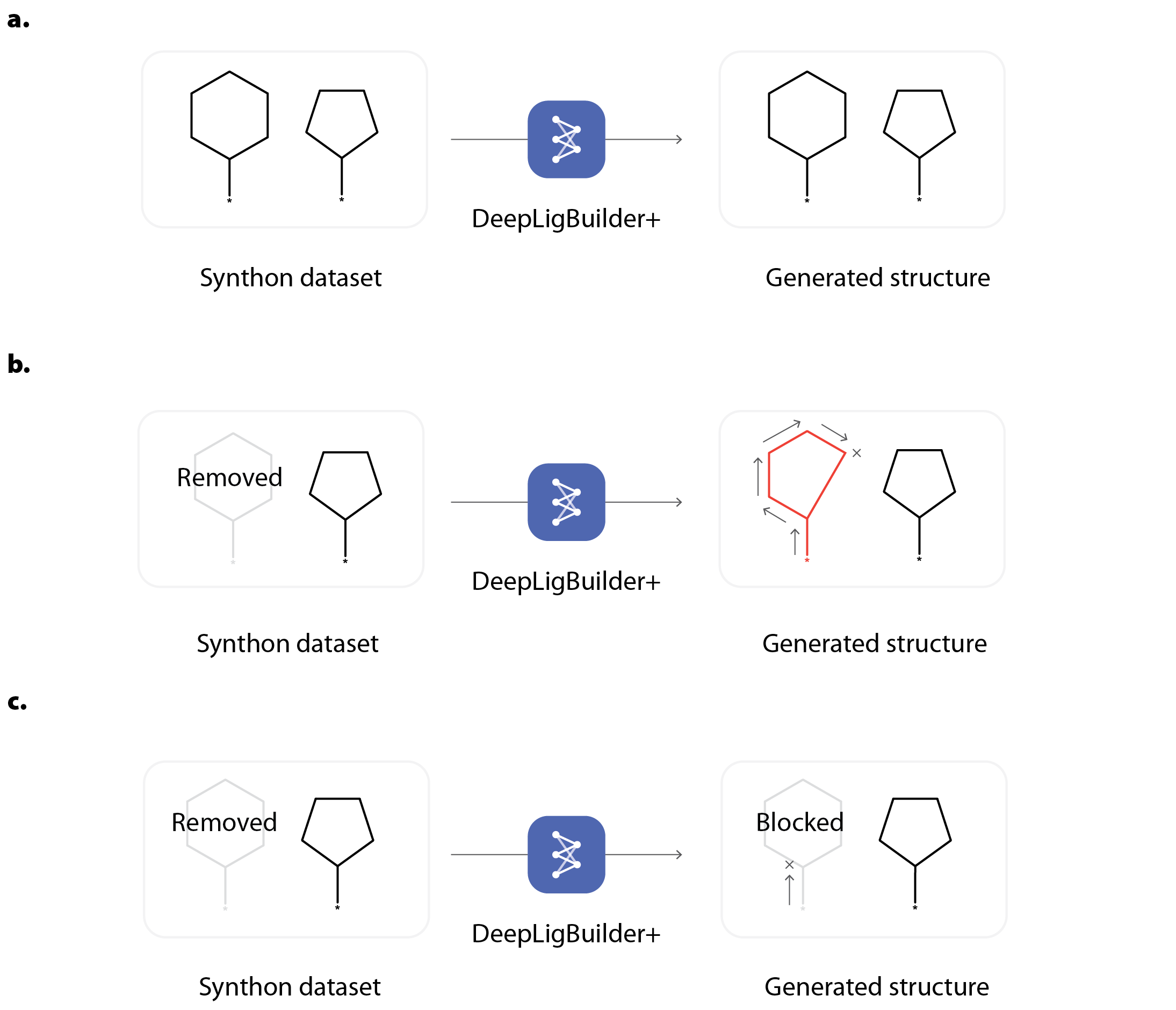}
\caption{Issues related to ring generation when the synthon dataset is
changed, and how the refined ring-based generation scheme can alleviate
this problem. \textbf{a.} The original synthon dataset, and the
structures of generated molecules. \textbf{b.} The six-membered ring is
removed from the dataset. Without retraining, the model will not
acknowledge this change. As a result, the model will still attempt to
generate six-membered rings, but the generation will be terminated in
advance due to the constraints on available synthons, resulting in
problematic structures shown below. \textbf{c.} However, if the ring
size is determined before generating its structures, the model will
acknowledge the change in the synthon dataset, since the action of
generating a six-membered ring is masked due to the imposed synthon
constraints, resulting in molecule structures with higher
quality.}\label{fig:s-ring-generation}
}
\end{figure}

A major difference between the generation scheme proposed in this work
compared with the previous version of DeepLigBuilder\cite{Li.2021}
is its emphasis on ring generation. Before generating explicit ring
structures, DeepLigBuilder+ will first determine the size of the ring,
as well as the location the ring will be closed. This information will
help to guide the process of ring generation. It will also help to avoid
problems when the user changes the synthon dataset (to be discussed in
the next section, also see Figure \ref{fig:s-ring-generation}).

\hypertarget{sec:s-prefix-tree}{%
\subsection{Constraining the model to generate structures inside the
synthon database}\label{sec:s-prefix-tree}}

To ensure synthetic accessibility, the synthons generated by
DeepLigBuilder+ must be restricted to the synthon dataset derived from
purchasable building blocks. To achieve this goal, most previous methods
use reactants as basic units for molecule generation and apply neural
networks to parametrize scoring functions that filter the reactant
dataset for appropriate candidates at each generation step. In this
work, we propose a radically different approach. Instead of adopting a
generation scheme based on reactants, we still use atoms as basic
generation units, building on the foundation of previous
works\cite{Li.2018,Li.2021}. To enforce constraints on the chemical
space, we apply masks on the action space at each step of generation, so
that we can ensure that the output topological structure can be found in
the synthon dataset.

Next, we show how such action masks can be calculated at each step.
First, for each synthon in the synthon database \(s \in \mathcal{S}\),
we represent it as a list of actions that can be used to generate its
molecular graph, which is indicated as \((a_1, ..., a_T)\rightarrow s\).
The definition of the action space and the generation process follows
Section S\ref{sec:s-mol-gen}. Since we only need to constrain the
topological structures, 3D action information, such as the position of
new atoms, is removed from the action space. For convenience, we refer
to such action sequences as trajectories and write them as \(\tau\). In
this way, we convert the synthon dataset into a collection of
trajectories
\(\mathcal{T}(\mathcal{S}) = \{\tau|\exists s\in \mathcal{S}\ \mathrm{s.t.\ } \tau \rightarrow s\}\).

To ensure synthons generated by the model lie in \(\mathcal{S}\), we
need to ensure that the generation trajectories lie in
\(\mathcal{T}(S)\). At each step \(t\), given the generation history
\(\tau_t=\tau[1..t-1]=(a_1,...,a_{t-1})\), in order to ensure that the
final trajectory will lie inside \(\mathcal{T}(\mathcal{S})\), we need
to make sure that the next action \(a_t\) is inside the following set:
\[
a_t \in \mathcal{A}(\tau_t, \mathcal{S})=\{a|\exists\tau'\in\mathcal{T}(\mathcal{S})\ \mathrm{s.t.}\ \tau'[1..t]=\tau_t\cdot a\}
\] It is easy to prove that as long as this requirement holds at each
step, we can guarantee that the resulting synthon will be inside
\(\mathcal{S}\). In order to efficiently calculate
\(\mathcal{A}(\tau_t, \mathcal{S})\), the trajectories in
\(\mathcal{T}(S)\) are organized into a prefix tree. Inside the tree,
each node represents a prefix of some trajectory in \(\mathcal{T}(S)\),
and each edge represents an action. During retrieval, we descend the
tree to find the node that equals \(\tau_t\), and then collect all its
outgoing edges. It can be seen that those edges form the set
\(\mathcal{A}(\tau_t, \mathcal{S})\). In this way, we can construct
action masks at each step to ensure that the resulting synthon can be
found in the synthon dataset.

As a method to constrain the chemical space of the generative model, our
method has some advantages compared with previous methods:

\begin{enumerate}
\def\labelenumi{(\arabic{enumi})}
\item
  The complexity of generating each synthon does not scale with the size
  of the synthon dataset. Searching inside the prefix tree has an
  average cost of \(O(L)\), where \(L\) is the average number of steps
  required to generate a synthon. Other approaches generally require a
  full scan of the reactant dataset, which has more limited scalability
  for larger synthon databases. Constructing the prefix tree will cost
  \(O(LN)\), where \(N\) is the size of the synthon, but we only need to
  construct the tree once, and it can then be reused for subsequent
  generation tasks.
\item
  Our method only constrains the 2D (topological) structure of
  molecules, while the 3D coordinates are generated by the neural
  network. This eliminates the need of building a 3D fragment database,
  as done in several previous works\cite{Powers.2022,Adams.2022}.
  Such approaches may cause some technical issues. First, enumerating 3D
  conformers will significantly increase the size of the fragment
  dataset, especially when the dataset contains large flexible
  structures. Second, the conformation of a fragment depends on its
  environment, and enumerating its conformation in isolation may result
  in inaccurate results when the fragment is attached to another
  molecule.
\end{enumerate}

Some practical issues need to be considered when applying this method:

\begin{enumerate}
\def\labelenumi{(\arabic{enumi})}
\item
  There are generally multiple ways to generate a molecular structure.
  To reduce complexity and computational cost, we follow the approach in
  previous works \cite{Li.2018,Li.2021}, which uses a depth-first,
  canonically ordered way to generate molecules. In this way, a
  molecular graph will correspond to exactly one trajectory, when the
  starting atom for generation is given. To make the model more
  flexible, we allow multiple starting points for the generation as in
  our previous work\cite{Li.2021}.
\item
  When using the proposed method to constrain the chemical space, there
  may be issues related to ring conformation when the synthon dataset is
  changed without model retraining. An illustrative example is given in
  Figure \ref{fig:s-ring-generation}. To alleviate this problem, we
  adopt a more refined ring-generation scheme, which allocates the size
  and location of the ring before its structures are generated, as
  discussed in the previous section.
\end{enumerate}

\hypertarget{sec:s-network}{%
\subsection{Network architecture}\label{sec:s-network}}

In this section, we give a detailed description of the neural network
architecture. We first describe the inputs required by the network.
Then, we show how the inputs are embedded before feeding to the neural
network. Next, we detail the architecture of the neural network.
Finally, we demonstrate how the output features from the transformer are
used to generate the action at each step with a MADE-based policy
network.

\subsubsection{Input features}

The model receives previously generated synthons as inputs, as well as
shape and pharmacophore information for structure-based generation
tasks.

\paragraph{Graph inputs}

All graph-related inputs, including previously generated synthons and
the intermediate synthon structures, are represented as their generation
trajectories \(\tau\). This acts as a sequence-based representation of
molecular graphs, similar to the concept of a ``sentence'' in NLP tasks.
Each action in the sequence corresponds to a ``token'' or ``word'' in
the sentence. The following information is included in each token:

\begin{enumerate}
\def\labelenumi{(\arabic{enumi})}
\item
  The current step id (\(t\));
\item
  Action performed at this step, including the action type (\(act_t\))
  and the type of new bonds added (\(nbt_t\));
\item
  Information of the focused node after the action is applied, such as
  its index (\(id_t\), ordered based on the step each atom is
  generated), element type (\(el_t\)), formal charge(\(fc_t\)), and the
  number of explicit hydrogens (\(neh_t\)) attached. Information about
  whether the node allows for branching (\(b_t\)) or whether the node
  can act as a target for ring closure is also included(\(r_t\)).
\item
  If a ring is being generated, we also input the expected size of the
  ring (\(rs_t\)) and the expected target for ring closure (\(rt_t\)).
\end{enumerate}

In addition to the information above, each token is attached with a 3D
coordinate frame \((\mathbf{o}_t, \mathbf{R}_t)\) using the method
developed in our previous work \cite{Li.2021}. Those frames have
several functionalities.

\begin{enumerate}
\def\labelenumi{(\arabic{enumi})}
\item
  The coordinates of newly generated atoms are defined under those
  frames. The spherical coordinate values of the local frames correspond
  to bond lengths, bond angles, and torsion angles.
\item
  Those frames are used in the IPA modules to communicate 3D information
  between the tokens.
\item
  Those frames are used to define the relative 3D positional embeddings
  between tokens (to be discussed below).
\end{enumerate}

Besides input features for each token, we also include features for each
action pair as relative embedding to increase the performance of the
model. Those pair features include:

\begin{enumerate}
\def\labelenumi{(\arabic{enumi})}
\item
  The topological distance between focused atoms at each step
  (\(top_{tt'}\)).
\item
  The distance between the focused atoms in the spanning tree induced by
  the generation trajectory (\(tree_{tt'}\), recall descriptions in
  Section S\ref{sec:s-mol-gen}).
\item
  The relative 3D positions between coordinate frames attached to each
  action (\(\mathbf{\Delta x}_{tt'}\)). Specifically, for the action
  pair \((a_t, a_{t'})\), we first calculate the displacement between
  the origins of each frame and then transform the coordinate values to
  the local coordinate frame attached to \(a_i\).
\end{enumerate}

\paragraph{Pharmacophore inputs}

The input pharmacophore model can be represented as a sequence of
individual pharmacophore features, with definitions adopted from
Align-it\cite{Taminau.2008}. Each pharmacophore \(p\) contains
information about its type (\(pt_p\)) and radius (\(pr_p\)). We use the
Euclidean distances (\(pd_{pp'}\)) between pharmacophore pairs
\((p, p')\) as relative positional embeddings.

In the decoder, we need to communicate between the pharmacophore model
and the synthon structure. Therefore, we feed the model with the
following information about the 3D relationship between each
pharmacophore-action pair:

\begin{enumerate}
\def\labelenumi{(\arabic{enumi})}
\item
  The position of each pharmacophore in the local coordinate system
  attached to each action (\(\mathbf{\Delta x}_{pt}\));
\item
  The direction for each pharmacophore (only HBDs and HBAs) in the local
  coordinate systems attached to each action
  (\(\hat{\mathbf{n}}_{pt}\)).
\end{enumerate}

\paragraph{Shape inputs}

The shape input can originate from known active ligands or directly from
the target pocket using programs such as
PANTHER\cite{Kurkinen.2019}. In both cases, we represent the input
shape as a set of 3D spheres. Most previous models use 3D-CNN to encode
shape information\cite{Skalic.2019}. This method is not equivariant
and induces additional computational costs. In comparison,
DeepLigBuilder+ uses a more compact, SE(3) equivariant representation
for 3D shapes based on 3D Zernike coefficients. For an input shape
composed of 3D spheres, we can represent it using a 3D scalar function
following Grant et al.\cite{Grant.1996}: \[
f(\mathbf{x})=\sum_{i=1}^{N}p_i\exp(-\alpha_i|\mathbf{x} - \mathbf{x}_i|^2)
\] Where \(\mathbf{x}_i\) is the location of each sphere, and \(p_i\)
and \(\alpha_i\) are parameters related to the radius of each sphere. We
define the function in the coordinate frame placed in the center of the
spheres \(\mathbf{x}^c=\frac{1}{N}\sum_{i=1}^N{\mathbf{x}_i}\). The
function is then decomposed as: \[
f(\mathbf{x})=\sum_{nlm}{c_{nl}^mZ_{nl}^m(\mathbf{x})}
\] Where \(Z_{nl}^m(\mathbf{x})\) are 3D Zernike
polynomials\cite{Wang.20112w5}. In this work, we use a truncated
series with \(n \le 9\). Those functions are generalizations of Zernike
polynomials defined in 2D space and act as the orthogonal basis for 3D
functions (defined in a ball with radius 1). Additionally, those
coefficients \(\mathbf{c} = \{c_{nl}^m\}_{nlm}\) changes in an
equivariant manner when a rotation \(\mathbf{R} \in SO(3)\) is applied
to the function \(f\): \[
f'(\mathbf{x})=f(\mathbf{R}^{-1}\mathbf{x})=\sum_{nlm}\sum_{m'}R_{mm'}^lc_{nl}^{m'}Z_{nl}^m(\mathbf{x})
\] Where \(R_{mm'}^l\) are the matrices that can be used to ``rotate''
the coefficients of the function \(f\). Those matrices can be
efficiently calculated using the methods proposed by Ivanic et
al.\cite{Ivanic.1998}

The representation of the shape can then be written as
\((\mathbf{x}^c, \mathbf{c})\). When feeding into the model, we rotate
it into the local coordinate frames of each action
\((\mathbf{x}^c_t, \mathbf{c}_t)\), and concatenate it with other action
features. In other words, the shape information is used by the
transformer similar to a positional embedding for each token.

\subsubsection{Embedding layers}

Several types of embedding layers are used for the input information
discussed above, including:

\begin{enumerate}
\def\labelenumi{(\arabic{enumi})}
\item
  Lookup tables with trainable parameters. This type of layer is used to
  embed atom, bond, and pharmacophore types, as well as topological
  distances. The full list of inputs includes \(act_t\), \(nbt_t\),
  \(el_t\), \(fc_t\), \(neh_t\), \(b_t\), \(r_t\), \(rs_t\),
  \(top_{tt'}\), \(tree_{tt'}\), and \(pt_p\).
\item
  Positional embedding using sine and cosine
  functions\cite{Vaswani.2017}. This type of embedding is widely
  used in transformer models to embed position-related data. In this
  work, we use it to embed time, position, and distance-related
  information. The full list includes \(t\), \(idx_t\), \(rt_t\),
  \(\mathbf{\Delta x}_{tt'}\), \(pd_{pp'}\), \(\mathbf{\Delta x}_{tp}\)
  and \(\mathbf{x}^c_t\).
\item
  Some inputs with continuous representations are input to the model
  as-is. This includes: \(pr_p\), \(\hat{\mathbf{n}}_{pt}\), and
  \(\mathbf{c}_t\).
\end{enumerate}

After the inputs are embedded, for each token (action or pharmacophore)
and token pair, we concatenate all input information into a vector and
use a linear layer to project the inputs to a predefined dimension,
which is then used as transformer inputs. We write the size of the
dimension as \(F\) for each token and \(F'\) for each token pair. In
this work, we have \(F'=\frac{F}{2}\) and two values \(\{512,256\}\) are
experimented for \(F\).

\subsubsection{The transformer architecture}

The transformer is responsible for processing the input features to
generate a state embedding at each step. Later, this state embedding
will be used by the policy network for action sampling. The transformer
consists of multiple encoders responsible for processing different
inputs, and a decoder used to generate the state embedder. The decoder
and encoders are composed of transformer layers, each containing one or
more attention layers and a tokenwise dense layer. The attention and
dense layers are wrapped inside residue blocks.

\paragraph{Attention layers}

Two types of attention layers are used in this work. The first is the
widely used scaled dot-product attention(SDPA)\cite{Vaswani.2017},
with additional relative positional bias. Given the features of the
source sequence \(\{\mathbf{h}_i^s\}_{i=1}^{l_s}\) , the target sequence
\(\{\mathbf{h}_i^t\}_{i=1}^{l_t}\), and the source-target token pairs
\(\{\mathbf{h}_{ij}^p\}_{i=1,...,l_t}^{j=1,...,l_s}\), the attention
layer performs the following operations to calculate the output feature
\(\{\mathbf{h'}_i\}_{i=1}^{l_t}\) for each target token:

\[
[\mathbf{k}_j^h, \mathbf{v}_j^h]_{h=1}^H=\mathrm{Linear}_{kv}(\mathbf{h}_j^s)
,\
[\mathbf{q}_i^h]_{h=1}^H=\mathrm{Linear}_q(\mathbf{h}_i^t)
,\
[b_{ij}^h,\mathbf{z}_{ij}^h]_{h=1}^H=\mathrm{Linear}_p(\mathbf{h}_{ij}^p)
\] \[
a_{ij}^h =\mathrm{softmax}_j(\frac{1}{\sqrt{d}}\mathbf{q}_i^h\cdot \mathbf{k}_j^h + b_{ij}^h)
\] \[
\mathbf{h'}_i=\mathrm{Linear}_{out}([\sum_{j=1}^{l_s}{a_{ij}^h\mathbf{v}_j^h},\sum_{j=1}^{l_s}{a_{ij}^h\mathbf{z}_{ij}^h}]_{h=1}^H)
\] \[
\mathrm{where}\ i=1,...,l_t;\ j=1,...,l_s;\ h=1,...,H
\]

\(H\) denotes the number of attention heads, which is set to be 16 in
this work. \(d\) is the dimension of each query, key, or value vector,
which is set to be \(d=\frac{F}{H}\), the \([\cdot]\) operator
represents concatenation or unpacking respectively when it appears in
the right or left side of the equations.

The second type of attention is invariant point attention (IPA),
initially proposed in AlphaFold2\cite{Jumper.2021}. Similar to SDPA,
we first calculate vector-based queries for the target sequence, as well
as the keys and values for the source sequence: \[
[\mathbf{k}_j^h, \mathbf{v}_j^h]_{h=1}^H=\mathrm{Linear}_{kv}(\mathbf{h}_j^s)
,\
[\mathbf{q}_i^h]_{h=1}^H=\mathrm{Linear}_q(\mathbf{h}_i^t)
,\
[b_{ij}^h,\mathbf{z}_{ij}^h]_{h=1}^H=\mathrm{Linear}_p(\mathbf{h}_{ij}^p)
\] \[
\mathrm{where}\ i=1,...,l_t;\ j=1,...,l_s
\]

Different from SDPA, IPA also calculates keys, queries, and values based
on 3D points: \[
[\vec{k}_i^{ph}, \vec{v}_i^{ph}]_{p=1,...,P}^{h=1,...,H}=\mathrm{Linear}_{kv}^\mathrm{3D}(\mathbf{h}_i^s)
,\
[\vec{q}_j^{ph}]_{p=1,...,P}^{h=1,...,H}=\mathrm{Linear}_q^\mathrm{3D}(\mathbf{h}_j^t)
\] \[
\mathrm{where}\ i=1,...,l_t;\ j=1,...,l_s
\]

Where \(P\) is the number of points for each attention head and is set
to be 8 in this work. Those points are defined on the local coordinate
system attached to each token and can be transformed into the global
coordinate system as: \[
\vec{k'}_j^{ph}=\mathbf{R}_j^s\vec{k}_j^{ph}+\mathbf{o}_j^s
,\
\vec{v'}_j^{ph}=\mathbf{R}_j^s\vec{v}_j^{ph}+\mathbf{o}_j^s
,\
\vec{q'}_i^{ph}=\mathbf{R}_i^t\vec{q}_i^{ph}+\mathbf{o}_i^t
\] \[
\mathrm{where}\ i=1,...,l_t;\ j=1,...,l_s;\ h=1,...,H;\ p=1,...,P
\] Where \(\{(\mathbf{o}_i^t, \mathbf{R}_i^t)\}_{i=1}^{l_t}\) and
\(\{(\mathbf{o}_i^s, \mathbf{R}_i^s)\}_{i=1}^{l_s}\) are coordinate
frames attached to each source and target tokens. Attention maps are
then calculated as: \[
a_{ij}^h =\mathrm{softmax}_j(w_L(\frac{1}{\sqrt{d}}\mathbf{q}_i^h\cdot \mathbf{k}_j^h + b_{ij}^h -\frac{\gamma^hw_C}{2}\sum_{p=1}^P|\vec{q'}_i^{ph} - \vec{k'}_j^{ph}|^2))
\] \[
\mathrm{where}\ i=1,...,l_t;\ j=1,...,l_s;\ h=1,...,H
\] In which \(w_L=\sqrt{\frac{1}{3}}\) and \(w_C=\sqrt{\frac{2}{9P}}\).
The values are then used to calculate the output features: \[
\mathbf{f}_i^h=\sum_{j=1}^{l_s}{a_{ij}^h\mathbf{v}_j^h}
\] \[
\tilde{\mathbf{f}}_i^h=\sum_{j=1}^{l_s}{a_{ij}^h\mathbf{z}_{ij}^h}
\] \[
\vec{f}_i^{hp}=(\mathbf{R}_i^t)^{-1}(\sum_{j=1}^{l_s}{a_{ij}^h\vec{v}_{j}^{ph}}-\mathbf{o}_i^t)
\] \[
\mathbf{h'}_i=\mathrm{Linear}_\mathrm{out}([\mathbf{f}_i^h,\tilde{\mathbf{f}}_i^h,[\vec{f}_i^{ph}]_{p=1}^P]_{h=1}^H)
\] \[
\mathrm{where}\ i=1,...,l_t;\ h=1,...,H;\ p=1,...,P
\] Note that compared with the original implementation, we do not
include the norm of \(\vec{f}_i^{hp}\) in the input of the linear
projection. SDPA and IPA are used in different situations in the
transformer network. The encoder for previous synthons uses IPA. In the
decoder, self-attention layers and outer-attention layers with previous
synthons use IPA. The encoder for pharmacophores and the decoder
outer-attention layers with pharmacophores use SDPA.

\paragraph{The token-wise dense layers (MLP layers)}

MLP layers consist of two dense layers, each with a
normalization-activation-linear architecture. In this work, we use layer
normalization\cite{Ba.2016} and ELU\cite{Clevert.2015} as
activation units. The number of hidden features is set to be \(2F\).

\paragraph{Transformer layers}

The attention layers and MLP layers are composed of transformer layers
that are later stacked into the encoder and decoder networks. Each
transformer layer in the encoder consists of one attention layer and one
MLP layer. Each transformer layer in the decoder consists of two
attention layers, one for self-attention and the other for outer
attention, and an MLP layer. The attention and MLP layers are all
wrapped inside residual blocks.

\paragraph{Encoders, decoders, and the transformer network}

Multiple transformer layers are stacked to form the encoder and
decoders. For unconditional generation tasks, we use 6 blocks for the
encoder of previous synthons and 6 blocks for the decoder. For
structure-based generation tasks, the model also receives shape and
pharmacophore-based information, which uses 3 more encoder and decoder
layers for processing. A shallow configuration is also experimented with
in unconditional generation tasks, as a way to demonstrate the
performance using different network scales. In this configuration, the
encoder and the decoder each uses 3 blocks of transformer layers.

Two versions of geometric transformers are developed in this work. An
unconditional transformer is used to access the ability of this method
to generate drug-like, geometrically valid molecules with high
synthesizability. A conditional one, which receives user-provided
pharmacophores and shapes as extract inputs, is used as the rollout
policy to accelerate MCTS in SBDD problems.

\hypertarget{sec:policy-network}{%
\subsubsection{The policy network}\label{sec:policy-network}}

Using the state embedding generated by the transformer network, a policy
network is then applied to generate the action for the next step. Before
specifying the architecture of this network, we need to first define the
action space. Two types of decisions need to be made at each step of the
generation. The first one relates to the topological structure of the
molecule, including the type of action to be carried out, the type of
the new atom and bond, the size of the new ring, etc. The second one
relates to the 3D molecular structure, that is the position of the new
atom.

For topological decisions, we iterate through the synthon dataset to
collect the actions that are needed to produce all the synthon
structures. This result in the topological action space
\(\mathcal{A}^\mathrm{topo}\). For 3D actions, we write the location of
the new atom added to each step in the local spherical coordinate system
attached to the focused node \((r, \theta,\phi)\). We then discretize
\(r\), \(\theta\) and \(\phi\), each using two integers. Take the
\(\phi\) coordinate for example. We first split its domain
\((-\pi,\pi]\) into \(N_1\) equal-sized intervals
\((-\pi+\frac{2\pi}{N_1}i,-\pi +\frac{2\pi}{N_1}(i+1)];i=1,...,N_1\),
and find the one containing the coordinate value \(\phi\). The interval
found is named \(\phi^\mathrm{crude}\). To achieve further precision,
\(\phi^\mathrm{crude}\) is then divided into \(N_2\) smaller chunks. We
find the one containing \(\phi\), and name it \(\phi^\mathrm{refined}\).
Similar procedures are applied for \(r\) and \(\theta\).

Following the definition above, we can now write the action as: \[
a = (a^\mathrm{topo}, r^\mathrm{crude}, r^\mathrm{refined}, \theta^\mathrm{crude}, \theta^\mathrm{refined}, \phi^\mathrm{crude}, \phi^\mathrm{refined})
\] \[
\mathrm{where}\ a^\mathrm{topo}\in\mathcal{A}^\mathrm{topo};
\] \[
r^\mathrm{crude},\theta^\mathrm{crude},\phi^\mathrm{crude} \in \{1,...,N_1\}
\] \[
r^\mathrm{refined},\theta^\mathrm{refined},\phi^\mathrm{refined} \in \{1,...,N_2\}
\] In this work, \(N_1\) is set to be 30 and \(N_2\) to be 32. The task
of the policy network is to parametrize the distribution of \(a\) using
the neural network: \(p_\mathbf{\eta}(a|\mathbf{h})\), where
\(\mathbf{\eta}\) is the parameter of the network, and \(\mathbf{h}\) is
the state embedding. To efficiently model the joint distribution of
discrete variables in \(a\), we factorize \(p_\mathbf{\eta}\)
autoregressively and use MADE (masked autoencoder for density
estimation) as the model architecture. The network contains 3 layers and
630 hidden units for each layer.

The general idea of the policy network is similar to the previous
version of DeepLigBuilder\cite{Li.2021}. The major difference is
that we now use a discretized action space for the 3D positioning of new
atoms. Previously, we found that modeling the continuous distribution of
atom positions faces numerical issues, and proposed SoftMADE to address
those issues. However, SoftMADE works by adding noise to the 3D
coordinates, which reduces the accuracy of the model. Here in
DeepLigBuilder+, we use a two-step discretization process, which ensures
the precision of the distribution and can also avoid numerical
instabilities in continuous distributions.

\hypertarget{sec:s-dataset}{%
\subsection{Dataset and network training}\label{sec:s-dataset}}

\subsubsection{Training the unconditional model}

We use a dataset containing drug-like molecules randomly assembled using
the building blocks in the Enamine global stock as the training set. The
assembling process follows a step-wise procedure, which is initialized
with a random building block sampled from the dataset. At each step, the
possible reactions that can happen to the molecule are enumerated. The
reaction type is then randomly selected from the results, and the next
reactant is sampled from the building block set based on the selected
reaction type. Molecules are assembled with three reactants combined
using two reaction steps. Several filters are applied to obtain
drug-like molecules, including (1) Lipinski's rule-of-5
(Ro5)\cite{Lipinski.2001}, (2) Veber's rule\cite{Veber.2002},
(3) PAINS patterns\cite{Baell.2014}, and (4) a
QED\cite{Bickerton.2012} threshold of 0.5. After this process, we
obtained approximately 1 million (974,917) molecules, with 4/5 of which
used as the training set, and the rest used for validation and testing.
The 3D conformers of those molecules are generated using RDKit, by first
using ETKDG to embed the molecules into 3D space, and then optimizing
them using MMFF94s. To create more stable conformers, at most 10
conformers are generated for each molecule, and the one with the lowest
energy is used for model training. Finally, those molecular structures
are converted to synthons and subsequently transformed into generation
trajectories to train the unconditional transformer.

The network is implemented using PyTorch, and Adam is used for model
optimization\cite{Kingma.2014}, with a linear learning rate warm-up
of one epoch to 0.001, followed by an exponential learning rate decay.
The decay rate is 0.01, and several decay frequencies are experimented
with (see Table \ref{tbl:s-hyperparam}). The batch size is set to 1024,
and the model is trained for 100 epochs using 4 A100 GPUs, which may
take 1-2 days to complete.

To train the shape and pharmacophore-conditioned model, a dataset of
input-output pairs is constructed. First, we extracted a set of
ligand-based pharmacophore models and shapes from the 3D ligands in the
PDBBind 2020 dataset\cite{Wang.2004}. We use PDBBind as the data
source due to its ligand diversity. It not only contains drug-like
ligands, but also metabolites, peptides, fragments, and other types of
ligands that lack drug-likeness but are still frequently used in
pharmacophore extraction and interaction analysis. In this way, we can
increase the diversity of the pharmacophore and shape inputs. Note that
we do not use the protein structure and bioactivity values inside the
PDBBind dataset, therefore the extracted information is fully unlabeled.
Future research may also consider that information to improve the
quality of the extracted pharmacophores.

An overall 12,456 pharmacophores and shapes are extracted. Next, we
align the assembled molecules to the extracted pharmacophores and filter
those with a good match to form the training set. This process requires
\(974,917\times12,456\) 3D alignment operations, and due to its time
cost, we utilize an approach based on sequential filtering:

\begin{enumerate}
\def\labelenumi{(\arabic{enumi})}
\item
  First, we perform similarity calculation based on USRCAT fingerprints,
  and filter the top 10,000 most similar molecules to each pharmacophore
  and shape query. Since USRCAT similarity does not involve 3D
  alignment, it can be carried out with high efficiency inside GPU;
\item
  Next, we perform 3D alignments between the \(10,000\times 12,456\)
  pairs of molecules and queries based on 3D shape. The shapes are
  represented as a combination of 3D Gaussian functions, as described
  previously\cite{Grant.1996}, each ``colored'' with a pharmacophore
  type assigned using a set of SMARTS patterns defined in RDKit. PCA is
  then performed on the point sets to obtain the principle axes, and
  those axes are aligned to form the initial pose. 4 candidate poses are
  created by rotating 180\(\degree\) around each axis. A gradient-based
  optimization process is then used to tune the rotation and translation
  to achieve the best overlap between two shapes. The alignment process
  is accomplished using an in-house PyTorch program. Shape similarity is
  computed using the aligned pose, and the top 100 most similar
  molecules are retained for each pharmacophore-shape query.
\item
  After shape-based alignment and filtering, a more refined
  pharmacophore-based filtering is used to further enrich molecule-query
  pairs with a good match. At this step, we retain the top 10 most
  similar molecules for each pharmacophore and shape-based query.
\end{enumerate}

The filtering process described above creates a dataset containing
\(10 \times 12,456=124,560\) input-output pairs for model training. When
training the conditional model, we use the pre-trained unconditional
model as the base model and add pharmacophore and shape-related layers
at the tail of the transformer. To avoid overtraining, only the
parameters of the newly added layers are allowed to change. The training
is performed for 160 epochs and the learning rate decay is performed for
every 30 steps. Other hyperparameters for training the conditional model
are similar to that used to train its unconditional counterpart.

\hypertarget{sec:s-mcts}{%
\subsection{Monte Carlo tree search}\label{sec:s-mcts}}

Monte Carlo tree search is a widely used technique in reinforcement
learning which finds promising solutions for a given problem by
strategically expanding the search tree. In this work, we combine MCTS
with the pharmacophore and shape-conditioned transformer for the design
of synthesizable 3D molecules inside a given pocket. In order to search
for promising molecular structures, MCTS maintains a look-ahead tree
\(\mathcal{T}\) and iteratively builds \(\mathcal{T}\) using four steps:
selection, expansion, simulation, and backpropagation. DeepLigBuilder+
uses a variant of MCTS that includes several modifications to better
suit it to the 3D molecule generation tasks. In this section, we first
describe the data structure of the look-ahead tree \(\mathcal{T}\), then
discuss how the tree is updated at each step, and finally specify the
details of the hyperparameters used during MCTS runs.

The look-ahead tree in MCTS is used to store the history of previous
visits, with each node representing an intermediate state during
molecule generation, and each edge representing an action carried out at
each step. In DeepLigBuilder+, we introduced several custom
modifications in the data structure of nodes and edges:

\begin{enumerate}
\def\labelenumi{(\arabic{enumi})}
\item
  Edges in the tree contain topological and 3D actions applied to the
  molecular graph at each step. The 3D action is discretized using the
  method introduced in Section \ref{sec:policy-network}. Note that the
  edge only stores the value of the \(\phi\) coordinate, or torsion
  angle. This is because the bond lengths \(r\) and bond angles
  \(\theta\) are largely determined by the bond types and ring sizes,
  which are already stored in the topological actions.
\item
  The torsion angles of new atoms are stored in a coarse-grained form,
  that is \(\phi^\mathrm{crude}\). In this way, nodes in \(\mathcal{T}\)
  now represent sets of molecule structures with similar 3D
  conformations. This has a similar effect of clustering intermediate
  states using the torsion fingerprint.
\end{enumerate}

At each step of MCTS, the following operation are carried out
consecutively:

\begin{enumerate}
\def\labelenumi{(\arabic{enumi})}
\item
  The selection operation, which chooses a promising state from the tree
  based on its estimated value function. We follow
  MENTS\cite{xiao2019maximum} and use E2W (Empirical Exponential
  Weight) to generate the selection policy. As mentioned that each node
  represents a cluster of states with the same topological structure and
  similar 3D conformation. We randomly select one of the states from the
  cluster;
\item
  The expansion operation, which enumerates all possible actions that
  can be carried out given a state. In practice, we found that although
  the allowed action space for the 3D generative model is large,
  generally only a small subset of actions will be selected by the
  model. Based on this observation, we first perform multiple
  independent sampling of actions given the state selected using the
  transformer, and cluster the actions based on their topological action
  and torsion angle, as described previously. This procedure is similar
  to pruning branches in the search tree with a small probability of
  being chosen by the transformer;
\item
  The simulation operation, in which a full rollout is performed based
  on the selected state, and the results are evaluated using the Smina
  scoring function. Note that all new states created in the expansion
  operation are used to perform the rollout, which acts as a form of
  leaf-level parallelism\cite{Chaslot.2008}. Additionally, all
  subsequent actions generated at each rollout are added to the tree,
  which may help to reduce the instability during the tree search.
\end{enumerate}

Compared with the previous version of DeepLigBuilder, during the
rollout, we uses a shape and pharmacophore-conditioned generative model

When evaluating the generated outcome, DeepLigBuilder+ uses a soft
version of the Smina score as the reward function: \[
R(m)=\frac{\mathrm{softplus}(-S(m)) + \sum_{i=1}^3{\mathrm{softplus}(-S(s_i))}}{2}
\] Where \(m\) is generated molecule, \(s_i,\ i\in \{1,2,3\}\) are the
synthons fragments composed of the molecule, \(S(\cdot)\) is the Smina
score (evaluated directly without minimization), and
\(\mathrm{softplus}(x)=\ln(1+\exp(x))\). Adding \(\mathrm{softplus}\) to
the equation helps to reduce large penalties from the clashes with the
pocket, making the reward function softer.

\begin{enumerate}
\def\labelenumi{(\arabic{enumi})}
\setcounter{enumi}{3}
\tightlist
\item
  The backtracking operation, in which the calculated rewards are used
  to update the Q value estimates for each edge. Following MENTS, we use
  the soft-bellman backup as the operator to update the Q values.
\end{enumerate}

The number of rollout steps for each case study is set to 100. To fully
utilize GPU resources, tree-level, root-level, and leaf-level
parallelism are applied\cite{Chaslot.2008}. During selection, an
overall of 16 nodes is selected at once for simulation. Also, 20 trees
are constructed independently for each case study. During expansion, 32
actions are sampled from the full action space for clustering, and those
actions are all used to form updated states for rollout. There are two
major parameters controlling the balance between exploration and
exploitation in MENTS\cite{xiao2019maximum}, the temperature
parameter \(\tau\) and the exploration parameter \(\epsilon\) in E2W. In
this work, we have \(\tau=0.25\) and \(\epsilon=0.05\). The MCTS program
uses one NVIDIA 3070 graphics card with 1 CPU core.

\subsection{Model evaluation}

Several evaluations are performed to examine the performance of
DeepLigBuilder+ in different aspects.

\subsubsection{The unconditional generation tasks}

For the unconditional model, we investigate whether it can generate
drug-like molecules with high-quality 3D structures. To this end, a
dataset containing molecules randomly assembled from the building blocks
without drug-likeness filters is constructed as the target of
comparison, and the following evaluation metrics are employed:

\begin{enumerate}
\def\labelenumi{(\arabic{enumi})}
\item
  Metrics related to the molecular properties. For each molecule, a
  series of 2D or 3D properties are calculated, including molecular
  weight, LogP, QED, the number of hydrogen bond donors and acceptors,
  the number of rotatable bonds, total and polar solvent accessible
  surface areas, and the radius of gyration. The distributions of those
  properties are then compared between the generated molecules, the
  assembled molecules, and the test set molecules. To make the
  comparison, the mean and standard deviation values are calculated for
  each property in each dataset. We also use a metric calculated using
  the RMSD between mean and standard deviation values for a quantitative
  measurement: \(W=\sqrt{(\mu_1-\mu_2)^2 + (\sigma_1-\sigma_2)^2}\).
  Mathematically, this metric is equivalent to the Wasserstein distance
  between Gaussian approximations of two distributions.
\item
  For a more quantitative measurement of whether the model correctly
  constructed the drug-like chemical space of synthesizable molecules,
  we evaluate the MMD\cite{Gretton.2012} between generated molecules
  and test set molecules using 2D (morgan) and 3D (USRCAT) fingerprints.
  MMD, which stands for the maximum mean discrepancy, is a widely used
  technique for evaluating the differences between distributions, and
  are applied in several previous works for the evaluation of molecule
  generative models\cite{Li.2019,Li.2021}. MMD can be calculated as
  follows: \[
  \mathrm{MMD}_u^2(X,Y)=\frac{1}{m(m-1)}\sum_{i\ne j}k(x_i, x_j)+\frac{1}{n(n-1)}\sum_{i\ne j}{k(y_i,y_j)} - \frac{2}{mn}\sum_{ij}{k(x_i,y_j)}
  \] Where \(X=\{x_i\}_{i=1}^m\) and \(Y=\{y_i\}_{i=1}^n\) are two
  datasets to be compared. \(k\) is the kernel function based on either
  the Morgan or USRCAT fingerprint.
\item
  To evaluate the quality of 3D structures, we first examine the quality
  of local geometries of generated molecules. Similar to the previous
  work\cite{Li.2021}, we compare the distribution of torsion angles
  between generated molecules and the test set molecules. The
  environments are described using torsion SMARTS patterns by Schärfer
  et al.\cite{Scharfer.2013}. The difference between torsion angle
  distributions is quantized using MMD. We use the cosine values of
  torsion angle difference as kernel function when calculating the MMD.
\item
  To further access the quality of generated conformers by the model, we
  optimize each generated molecule using the MMFF94s force field and
  calculate the RMSD value between conformers before and after the
  optimization. To provide a context of the model's performance, we also
  perform this evaluation on conformers generated using the
  ETKDG\cite{Riniker.2015} method. This method is initially proposed
  as a faster alternative for forcefield-based conformation
  optimization.
\end{enumerate}

\subsubsection{The structure-based generation tasks}

When evaluating the performance of DeepLigBuilder+ in structure-based
generation tasks, we mainly focus on answering the following two
questions:

\begin{enumerate}
\def\labelenumi{(\arabic{enumi})}
\item
  Can MCTS help enrich molecules with high docking scores?
\item
  Can shape-based and pharmacophore-based conditional rollout policy
  help MCTS to discover better results faster?
\end{enumerate}

A series of ablation studies are performed to answer those two
questions. First, in terms of the benefit of MCTS-based sampling, we
compare the distribution of Smina docking scores for molecules generated
with or without MCTS. When calculating the Smina scores, we first move
the generated molecules out of the protein pocket, perform local
relaxation for each molecule using the MMFF94s forcefield, and then
re-dock the relaxed conformers back into the target pocket using Smina.

To access the benefit of the conditional rollout policy, the following
studies are performed:

\begin{enumerate}
\def\labelenumi{(\arabic{enumi})}
\item
  We determine whether the conditional model can achieve enrichment in
  pharmacophore and shape compared with its unconditional counterpart.
  This question can be answered by examining the distribution of shape
  and pharmacophore similarity between generated molecules and input
  queries.
\item
  We investigate whether the extra conditional inputs can help MCTS to
  achieve faster search speeds. To this end, ablation studies are
  performed by enabling and then disabling the conditional inputs for
  the rollout policy. At each MCTS step, we record the reward value for
  the best molecule found so far. We then compare whether the
  conditional rollout policy can help MCTS reach solutions with higher
  rewards faster.
\end{enumerate}

\section{Supplementary Figures}

\newpage

\newpage

\newpage

\newpage

\newpage

\begin{figure}
\hypertarget{fig:s-uncond-samples}{%
\centering
\includegraphics[width=1\textwidth,height=\textheight]{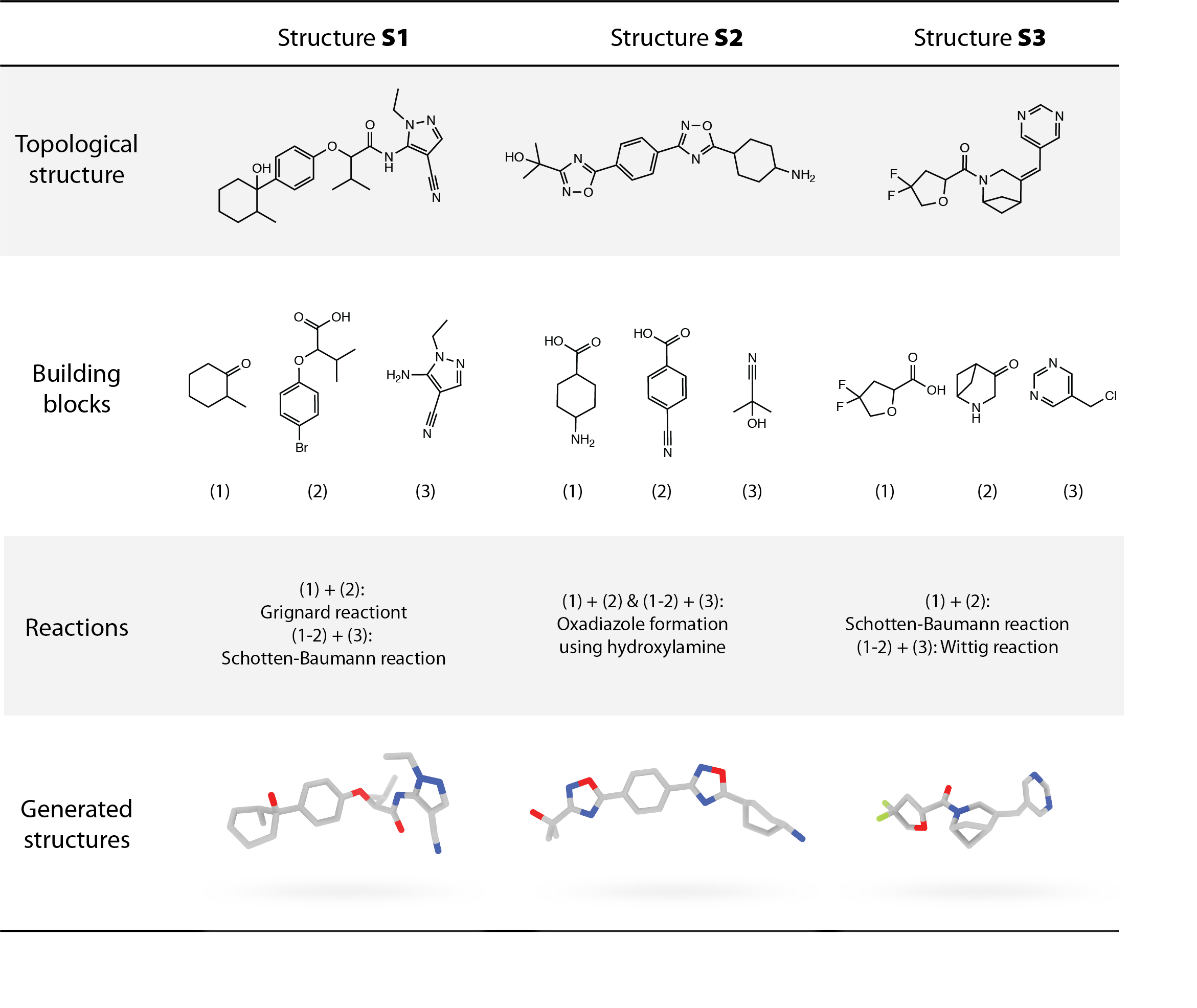}
\caption{Examples of several generated 3D molecules using the
unconditional transformer, along with the building blocks and proposed
reactions to synthesize the molecules.}\label{fig:s-uncond-samples}
}
\end{figure}

\newpage

\begin{figure}
\hypertarget{fig:s-prop}{%
\centering
\includegraphics[width=1\textwidth,height=\textheight]{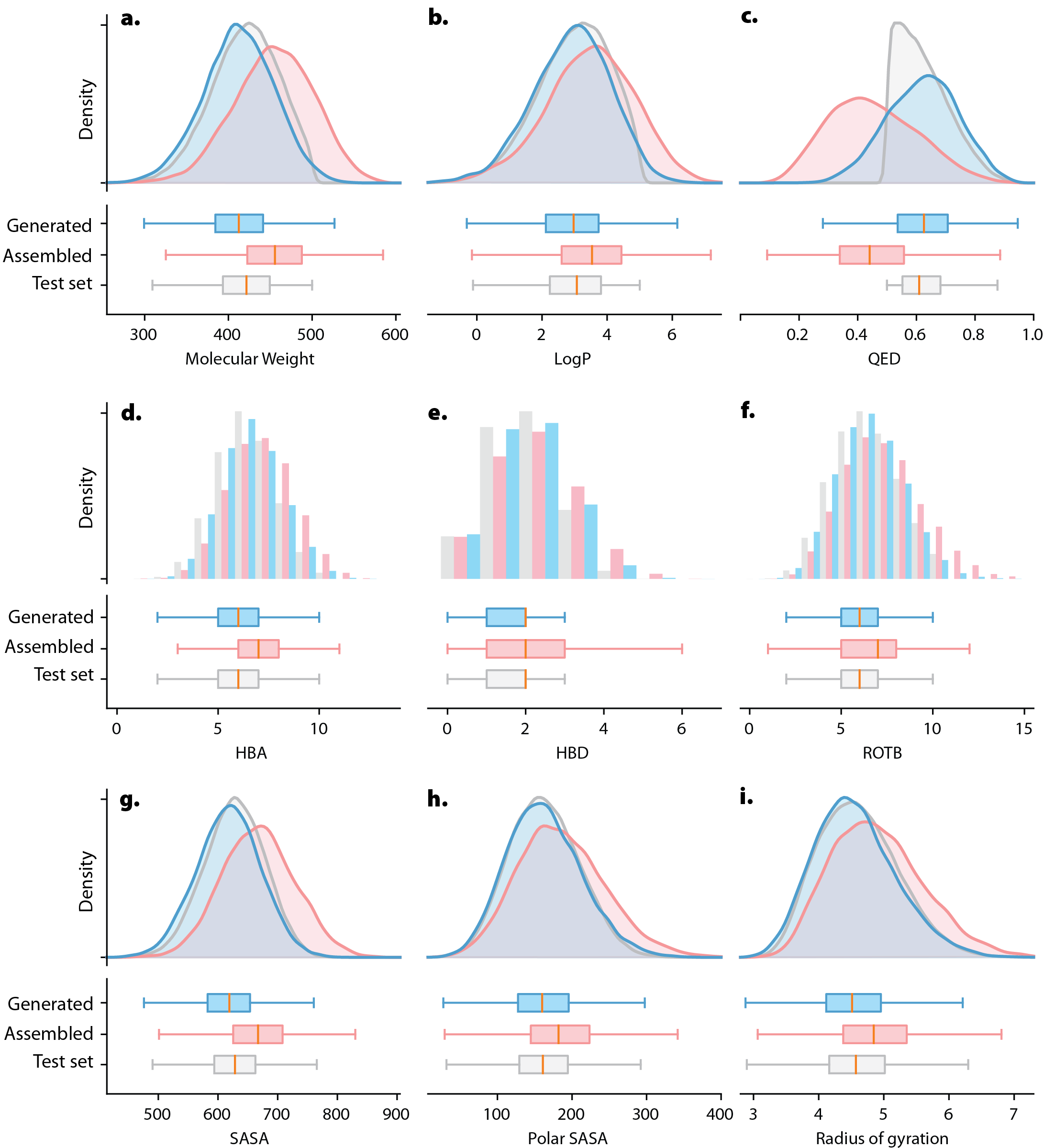}
\caption{The distribution of 2D and 3D molecular properties of the
generated (blue), test set (grey), and randomly assembled (red)
molecules. 2D properties includes: \textbf{a.} Molecular weight,
\textbf{b.} LogP, \textbf{c.} QED, \textbf{d.} the number of hydrogen
bond acceptors (HBA), \textbf{e.} the number of hydrogen bond donors,
\textbf{f.} The number of rotatable bonds (ROT). 3D properties includes:
\textbf{g.} The total amount of solvent accessible surface area (SASA),
\textbf{h.} polar solvent accessible surface areas (PolarSASA),
\textbf{i.} the radius of gyration.}\label{fig:s-prop}
}
\end{figure}

\newpage

\begin{figure}
\hypertarget{fig:s-tsne}{%
\centering
\includegraphics[width=1\textwidth,height=\textheight]{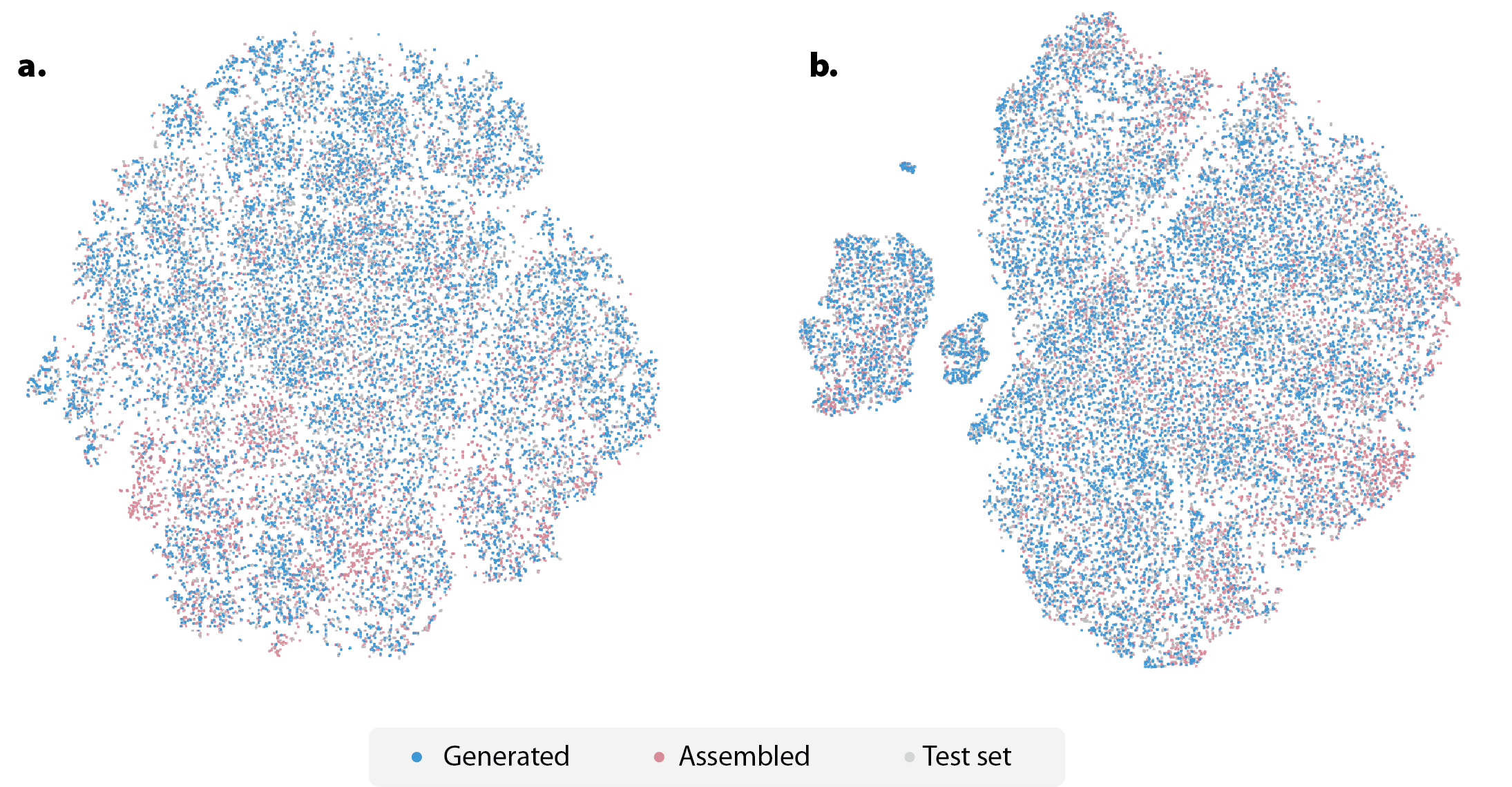}
\caption{A t-SNE visualization of the distribution of \textbf{a.} Morgan
fingerprint and \textbf{b.} USRCAT fingerprint. Blue dots represent
generated molecules, grey dots represent test set molecules, and red
dots represents molecules randomly assembled from the building
blocks.}\label{fig:s-tsne}
}
\end{figure}

\newpage

\begin{figure}
\hypertarget{fig:s-rmsd}{%
\centering
\includegraphics{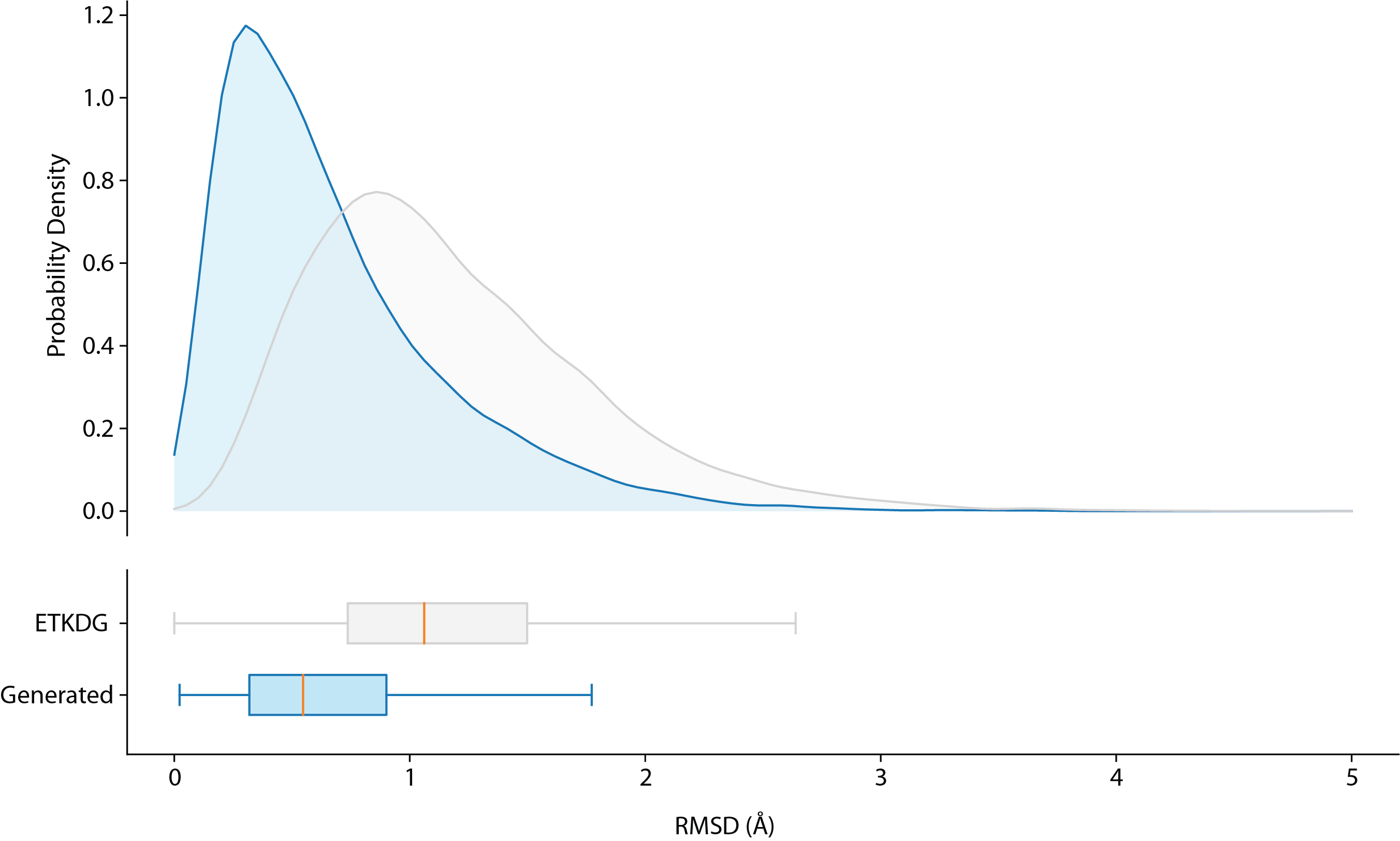}
\caption{The distribution of RMSD values after relaxation with MMFF94s.
Blue: molecules with conformations generated by the network. Grey:
molecules with conformations generated by the ETKDG method provided by
RDKit}\label{fig:s-rmsd}
}
\end{figure}

\newpage

\begin{figure}
\hypertarget{fig:s-mcts}{%
\centering
\includegraphics[width=1\textwidth,height=\textheight]{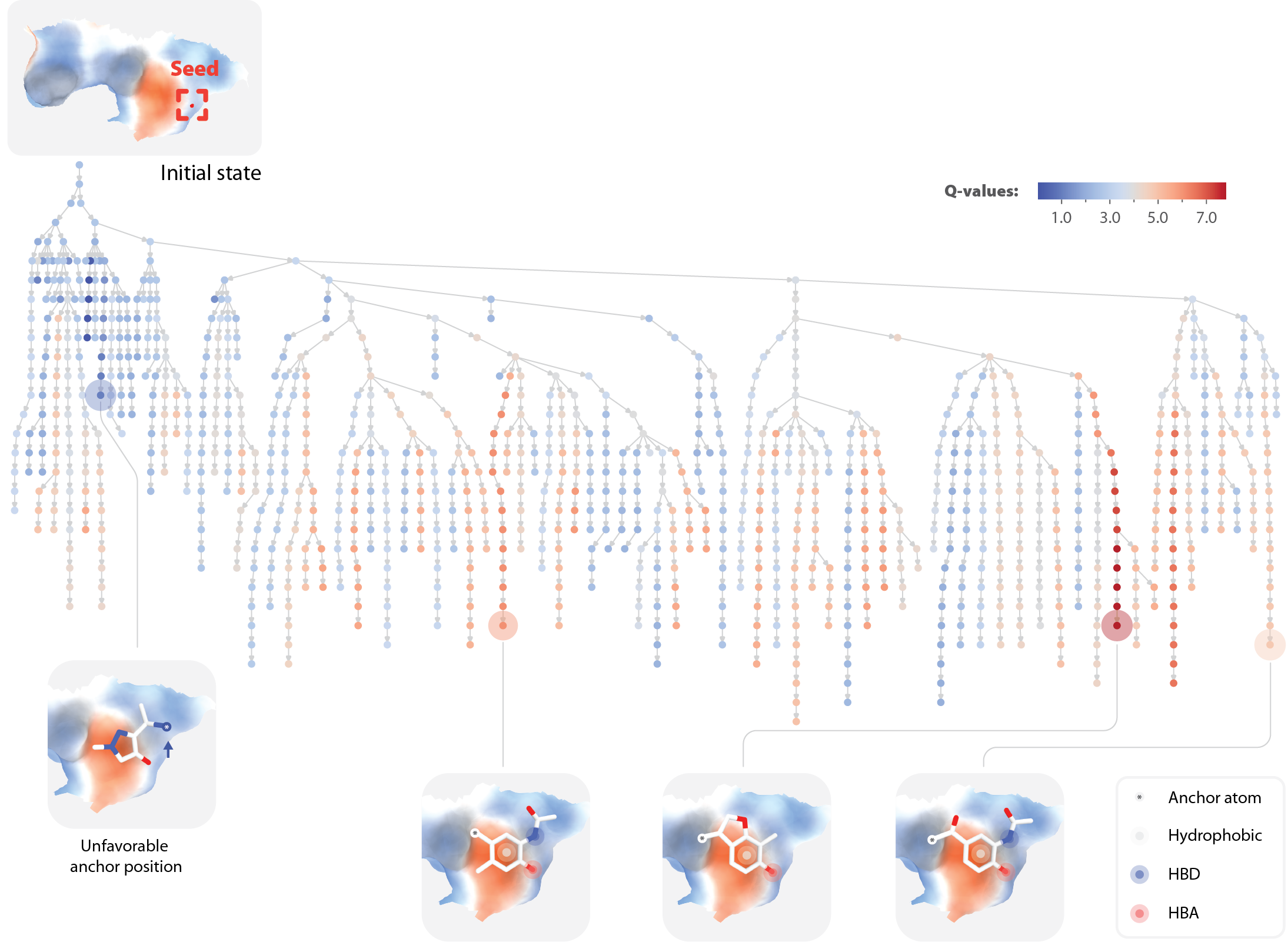}
\caption{The detailed structure of the search tree for the first synthon
in the BTK's ATP-binding pocket (only containing nodes with visit count
larger than 25). Some generated synthon structures are shown below the
tree, with important pharmacophore features highlighted. The leftmost
example shows a state with a low estimated Q-value, which largely
resulted from the unfavorable anchor position. As the result, this state
is not as frequently visited as other states with higher
Q-values.}\label{fig:s-mcts}
}
\end{figure}

\newpage

\begin{figure}
\hypertarget{fig:s-gdc-0853}{%
\centering
\includegraphics[width=1\textwidth,height=\textheight]{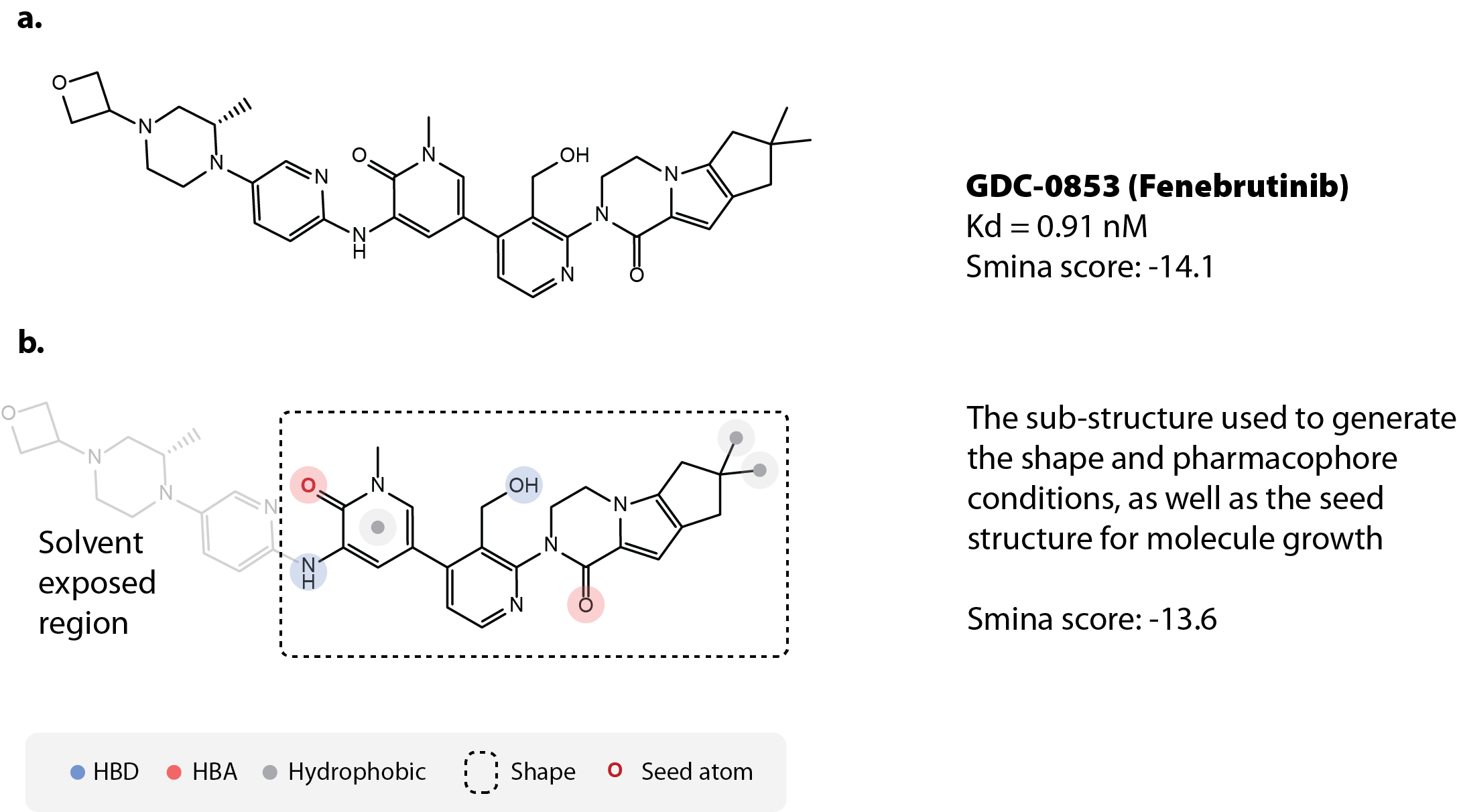}
\caption{\textbf{a.} The topological structure of GDC-0853
(Fenebrutinib). \textbf{b.} The part of GDC-0853 used extract
pharmacophore, shape, and seed for molecule generation. Note that some
part of GDC-0853 inside the solvent-exposed region is not considered for
pharmacophore and shape extraction.}\label{fig:s-gdc-0853}
}
\end{figure}

\newpage

\begin{figure}
\hypertarget{fig:s-btk-reaction}{%
\centering
\includegraphics[width=1\textwidth,height=\textheight]{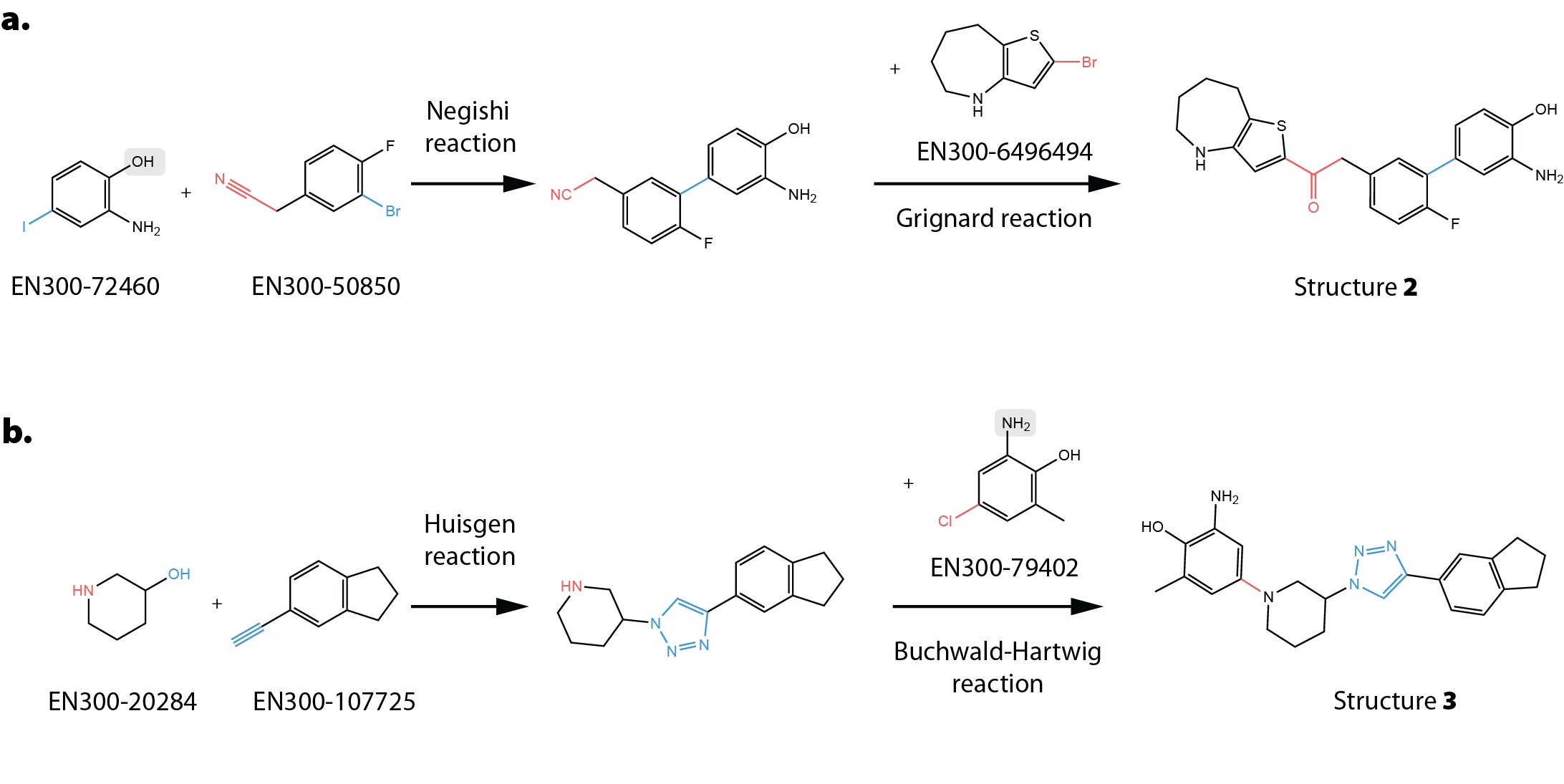}
\caption{The proposed synthetic path for Structure \textbf{2}
(\textbf{a}) and Structure \textbf{3} (\textbf{b}) by the model.
Functional groups marked in grey needs to be protected before the
reactions.}\label{fig:s-btk-reaction}
}
\end{figure}

\newpage

\begin{figure}
\hypertarget{fig:s-compound-15}{%
\centering
\includegraphics[width=1\textwidth,height=\textheight]{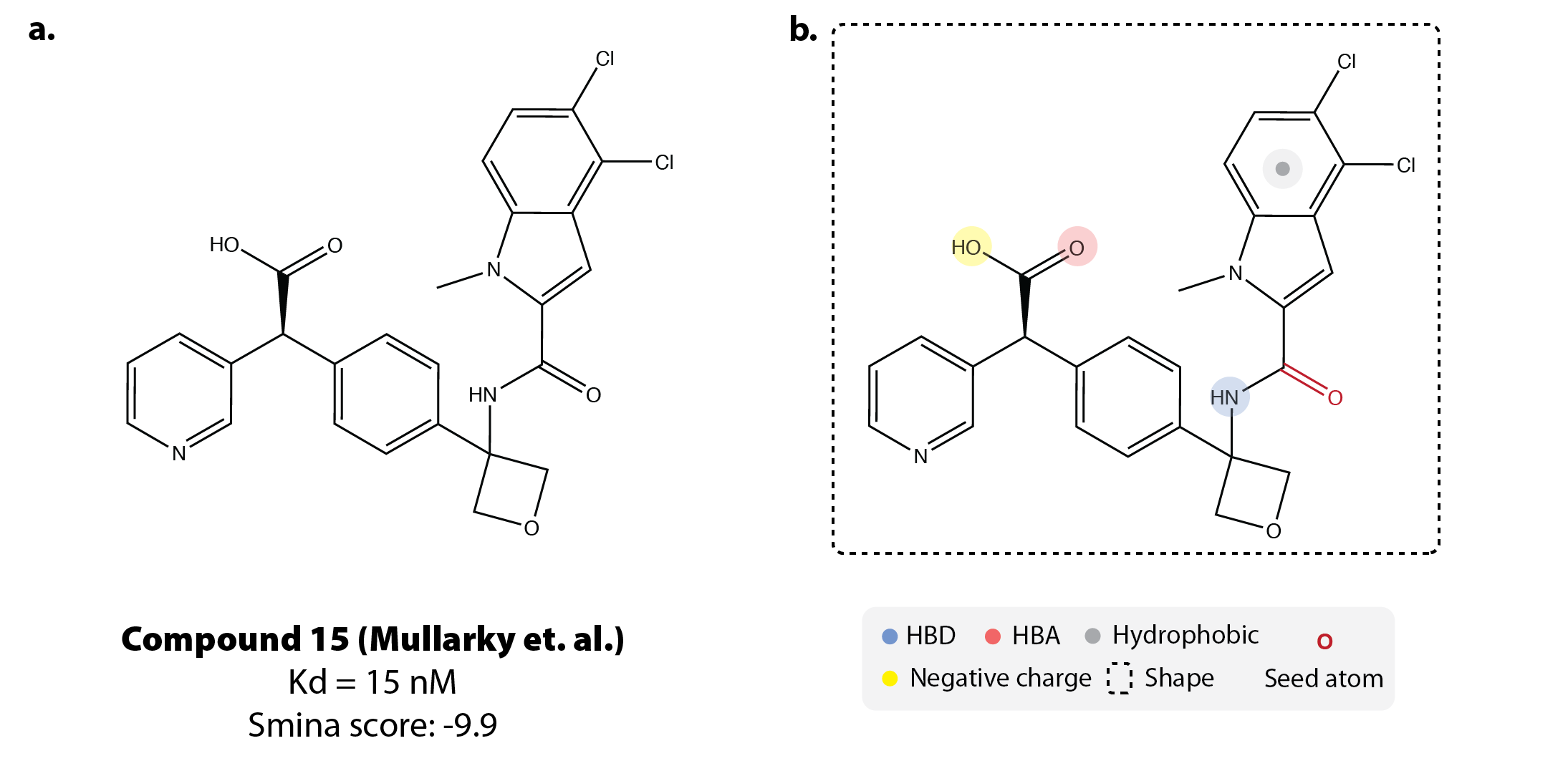}
\caption{\textbf{a.} The topological structure of Compound 15, a potent
inhibitor targeting the NAD pocket of PHGDH. \textbf{b.} The
pharmacophore features extracted from the binding mode of the molecule,
as well as the seed location for molecule
growth.}\label{fig:s-compound-15}
}
\end{figure}

\newpage

\begin{figure}
\hypertarget{fig:s-phgdh-reaction}{%
\centering
\includegraphics[width=1\textwidth,height=\textheight]{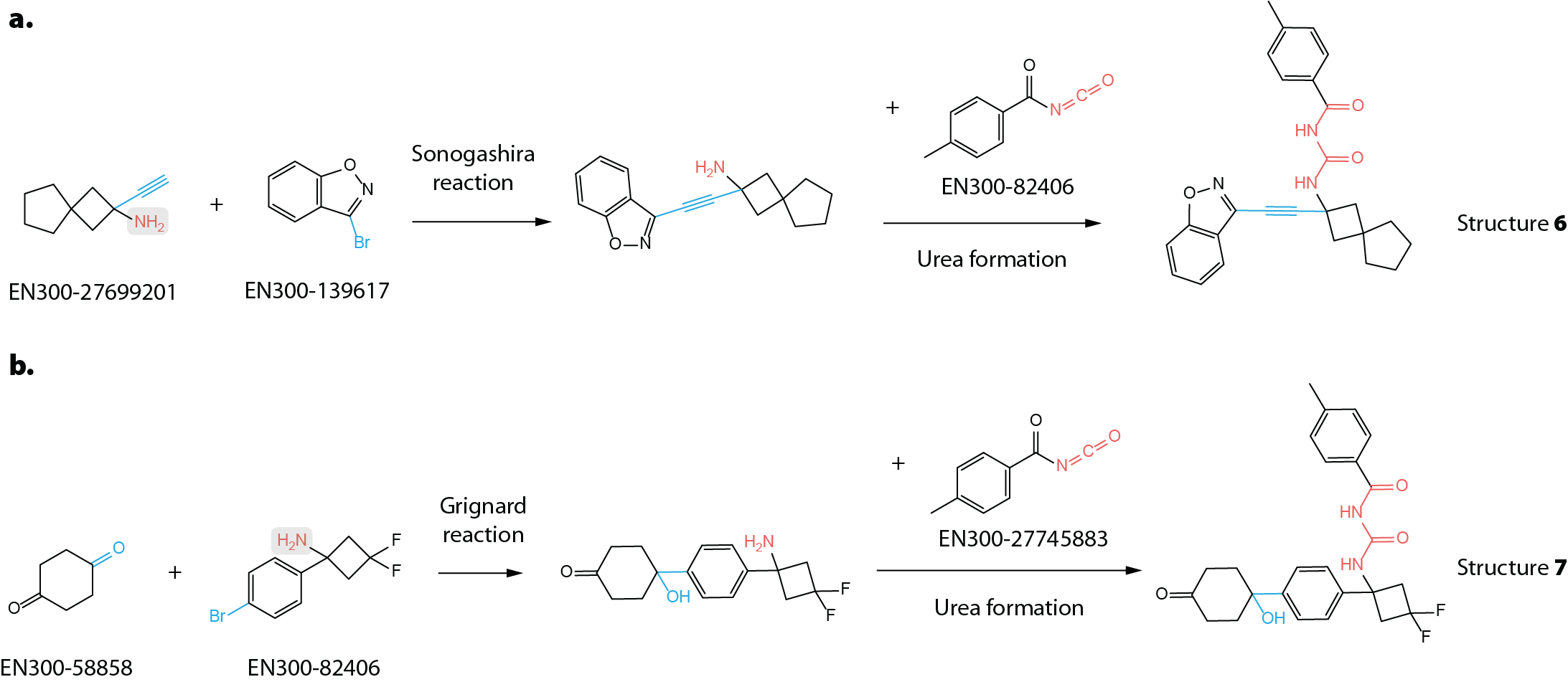}
\caption{The proposed synthetic path for Structure \textbf{6}
(\textbf{a}) and Structure \textbf{7} (\textbf{b}) by the model.
Functional groups marked in grey needs to be protected before the
reactions.}\label{fig:s-phgdh-reaction}
}
\end{figure}

\newpage

{\small
\begin{landscape}
\section{Supplementary tables}

\hypertarget{tbl:s-hyperparam}{}
\begin{longtable}[]{@{}
  >{\raggedright\arraybackslash}p{(\columnwidth - 18\tabcolsep) * \real{0.0663}}
  >{\raggedright\arraybackslash}p{(\columnwidth - 18\tabcolsep) * \real{0.1506}}
  >{\raggedright\arraybackslash}p{(\columnwidth - 18\tabcolsep) * \real{0.1084}}
  >{\raggedright\arraybackslash}p{(\columnwidth - 18\tabcolsep) * \real{0.1024}}
  >{\raggedright\arraybackslash}p{(\columnwidth - 18\tabcolsep) * \real{0.0663}}
  >{\raggedright\arraybackslash}p{(\columnwidth - 18\tabcolsep) * \real{0.0301}}
  >{\raggedright\arraybackslash}p{(\columnwidth - 18\tabcolsep) * \real{0.1446}}
  >{\raggedright\arraybackslash}p{(\columnwidth - 18\tabcolsep) * \real{0.1024}}
  >{\raggedright\arraybackslash}p{(\columnwidth - 18\tabcolsep) * \real{0.1084}}
  >{\raggedright\arraybackslash}p{(\columnwidth - 18\tabcolsep) * \real{0.1205}}@{}}
\caption{\label{tbl:s-hyperparam}A summary of different hyperparameter
configurations experimented in this work (BBs: building blocks, LR:
learning rate)}\tabularnewline
\toprule()
\begin{minipage}[b]{\linewidth}\raggedright
\end{minipage} & \begin{minipage}[b]{\linewidth}\raggedright
\end{minipage} & \begin{minipage}[b]{\linewidth}\raggedright
Model architecture
\end{minipage} & \begin{minipage}[b]{\linewidth}\raggedright
\end{minipage} & \begin{minipage}[b]{\linewidth}\raggedright
\end{minipage} & \begin{minipage}[b]{\linewidth}\raggedright
\end{minipage} & \begin{minipage}[b]{\linewidth}\raggedright
Training parameter
\end{minipage} & \begin{minipage}[b]{\linewidth}\raggedright
\end{minipage} & \begin{minipage}[b]{\linewidth}\raggedright
\end{minipage} & \begin{minipage}[b]{\linewidth}\raggedright
Generation parameter
\end{minipage} \\
\midrule()
\endfirsthead
\toprule()
\begin{minipage}[b]{\linewidth}\raggedright
\end{minipage} & \begin{minipage}[b]{\linewidth}\raggedright
\end{minipage} & \begin{minipage}[b]{\linewidth}\raggedright
Model architecture
\end{minipage} & \begin{minipage}[b]{\linewidth}\raggedright
\end{minipage} & \begin{minipage}[b]{\linewidth}\raggedright
\end{minipage} & \begin{minipage}[b]{\linewidth}\raggedright
\end{minipage} & \begin{minipage}[b]{\linewidth}\raggedright
Training parameter
\end{minipage} & \begin{minipage}[b]{\linewidth}\raggedright
\end{minipage} & \begin{minipage}[b]{\linewidth}\raggedright
\end{minipage} & \begin{minipage}[b]{\linewidth}\raggedright
Generation parameter
\end{minipage} \\
\midrule()
\endhead
Group & Variant name & IPA & 3D pair embedding & Width (\(F\)) & Depth &
Decay frequency (\#steps) & Noise probability & Noise scale(\(\text{\AA}\)) &
Building block \\
& Default configuration & \(\checkmark\) & \(\checkmark\) & 512 & 6 &
150 & 0.5 & 0.1 & Global \\
BBs & EU stock & \(\checkmark\) & \(\checkmark\) & 512 & 6 & 150 & 0.5 &
0.1 & \textbf{EU} \\
& Comprehensive catalog & \(\checkmark\) & \(\checkmark\) & 512 & 6 &
150 & 0.5 & 0.1 & \textbf{Comprehensive} \\
3D encoding & Dropped IPA & \(\times\) & \(\checkmark\) & 512 & 6 & 150
& 0.5 & 0.1 & Global \\
& Dropped 3D pair embedding & \(\checkmark\) & \(\times\) & 512 & 6 &
150 & 0.5 & 0.1 & Global \\
Scale & Narrow network & \(\checkmark\) & \(\checkmark\) & \textbf{256}
& 6 & 150 & 0.5 & 0.1 & Global \\
& Shallow network & \(\checkmark\) & \(\checkmark\) & 512 & \textbf{3} &
150 & 0.5 & 0.1 & Global \\
LR & Fast learning rate decay & \(\checkmark\) & \(\checkmark\) & 512 &
6 & \textbf{70} & 0.5 & 0.1 & Global \\
& Slow learning rate decay & \(\checkmark\) & \(\checkmark\) & 512 & 6 &
\textbf{300} & 0.5 & 0.1 & Global \\
Noise & High noise probability & \(\checkmark\) & \(\checkmark\) & 512 &
6 & 150 & \textbf{0.9} & 0.1 & Global \\
& Low noise probability & \(\checkmark\) & \(\checkmark\) & 512 & 6 &
150 & \textbf{0.1} & 0.1 & Global \\
& High noise scale & \(\checkmark\) & \(\checkmark\) & 512 & 6 & 150 &
0.5 & \textbf{0.2} & Global \\
& Low noise scale & \(\checkmark\) & \(\checkmark\) & 512 & 6 & 150 &
0.5 & \textbf{0.05} & Global \\
\bottomrule()
\end{longtable}

\newpage

\hypertarget{tbl:s-prop-2d}{}
\begin{longtable}[]{@{}
  >{\raggedright\arraybackslash}p{(\columnwidth - 36\tabcolsep) * \real{0.2451}}
  >{\raggedright\arraybackslash}p{(\columnwidth - 36\tabcolsep) * \real{0.0588}}
  >{\raggedright\arraybackslash}p{(\columnwidth - 36\tabcolsep) * \real{0.0490}}
  >{\raggedright\arraybackslash}p{(\columnwidth - 36\tabcolsep) * \real{0.0490}}
  >{\raggedright\arraybackslash}p{(\columnwidth - 36\tabcolsep) * \real{0.0392}}
  >{\raggedright\arraybackslash}p{(\columnwidth - 36\tabcolsep) * \real{0.0392}}
  >{\raggedright\arraybackslash}p{(\columnwidth - 36\tabcolsep) * \real{0.0392}}
  >{\raggedright\arraybackslash}p{(\columnwidth - 36\tabcolsep) * \real{0.0392}}
  >{\raggedright\arraybackslash}p{(\columnwidth - 36\tabcolsep) * \real{0.0392}}
  >{\raggedright\arraybackslash}p{(\columnwidth - 36\tabcolsep) * \real{0.0392}}
  >{\raggedright\arraybackslash}p{(\columnwidth - 36\tabcolsep) * \real{0.0392}}
  >{\raggedright\arraybackslash}p{(\columnwidth - 36\tabcolsep) * \real{0.0392}}
  >{\raggedright\arraybackslash}p{(\columnwidth - 36\tabcolsep) * \real{0.0392}}
  >{\raggedright\arraybackslash}p{(\columnwidth - 36\tabcolsep) * \real{0.0392}}
  >{\raggedright\arraybackslash}p{(\columnwidth - 36\tabcolsep) * \real{0.0392}}
  >{\raggedright\arraybackslash}p{(\columnwidth - 36\tabcolsep) * \real{0.0392}}
  >{\raggedright\arraybackslash}p{(\columnwidth - 36\tabcolsep) * \real{0.0392}}
  >{\raggedright\arraybackslash}p{(\columnwidth - 36\tabcolsep) * \real{0.0392}}
  >{\raggedright\arraybackslash}p{(\columnwidth - 36\tabcolsep) * \real{0.0490}}@{}}
\caption{\label{tbl:s-prop-2d}The distribution of 2D molecular
properties among molecules generated from models with different
hyperparameters (see Table \ref{tbl:s-hyperparam} for a detailed
description of each configuration). We report the mean and standard
deviation for each property, as well as the estimated Wasserstein
distance to the test set. For the first row, we report the statistics of
molecules randomly assembled from building blocks without any
drug-likeness filtering.}\tabularnewline
\toprule()
\begin{minipage}[b]{\linewidth}\raggedright
Model variant
\end{minipage} & \begin{minipage}[b]{\linewidth}\raggedright
MW
\end{minipage} & \begin{minipage}[b]{\linewidth}\raggedright
\end{minipage} & \begin{minipage}[b]{\linewidth}\raggedright
\end{minipage} & \begin{minipage}[b]{\linewidth}\raggedright
LogP
\end{minipage} & \begin{minipage}[b]{\linewidth}\raggedright
\end{minipage} & \begin{minipage}[b]{\linewidth}\raggedright
\end{minipage} & \begin{minipage}[b]{\linewidth}\raggedright
HBA
\end{minipage} & \begin{minipage}[b]{\linewidth}\raggedright
\end{minipage} & \begin{minipage}[b]{\linewidth}\raggedright
\end{minipage} & \begin{minipage}[b]{\linewidth}\raggedright
HBD
\end{minipage} & \begin{minipage}[b]{\linewidth}\raggedright
\end{minipage} & \begin{minipage}[b]{\linewidth}\raggedright
\end{minipage} & \begin{minipage}[b]{\linewidth}\raggedright
ROT
\end{minipage} & \begin{minipage}[b]{\linewidth}\raggedright
\end{minipage} & \begin{minipage}[b]{\linewidth}\raggedright
\end{minipage} & \begin{minipage}[b]{\linewidth}\raggedright
QED
\end{minipage} & \begin{minipage}[b]{\linewidth}\raggedright
\end{minipage} & \begin{minipage}[b]{\linewidth}\raggedright
\end{minipage} \\
\midrule()
\endfirsthead
\toprule()
\begin{minipage}[b]{\linewidth}\raggedright
Model variant
\end{minipage} & \begin{minipage}[b]{\linewidth}\raggedright
MW
\end{minipage} & \begin{minipage}[b]{\linewidth}\raggedright
\end{minipage} & \begin{minipage}[b]{\linewidth}\raggedright
\end{minipage} & \begin{minipage}[b]{\linewidth}\raggedright
LogP
\end{minipage} & \begin{minipage}[b]{\linewidth}\raggedright
\end{minipage} & \begin{minipage}[b]{\linewidth}\raggedright
\end{minipage} & \begin{minipage}[b]{\linewidth}\raggedright
HBA
\end{minipage} & \begin{minipage}[b]{\linewidth}\raggedright
\end{minipage} & \begin{minipage}[b]{\linewidth}\raggedright
\end{minipage} & \begin{minipage}[b]{\linewidth}\raggedright
HBD
\end{minipage} & \begin{minipage}[b]{\linewidth}\raggedright
\end{minipage} & \begin{minipage}[b]{\linewidth}\raggedright
\end{minipage} & \begin{minipage}[b]{\linewidth}\raggedright
ROT
\end{minipage} & \begin{minipage}[b]{\linewidth}\raggedright
\end{minipage} & \begin{minipage}[b]{\linewidth}\raggedright
\end{minipage} & \begin{minipage}[b]{\linewidth}\raggedright
QED
\end{minipage} & \begin{minipage}[b]{\linewidth}\raggedright
\end{minipage} & \begin{minipage}[b]{\linewidth}\raggedright
\end{minipage} \\
\midrule()
\endhead
& mean & std & wd & mean & std & wd & mean & std & wd & mean & std & wd
& mean & std & wd & mean & std & wd \\
Randomly assembled & 454.32 & 47.45 & 35.46 & 3.48 & 1.37 & 0.57 & 6.81
& 1.64 & 0.7 & 1.91 & 1.12 & 0.33 & 6.96 & 2.22 & 0.93 & 0.45 & 0.15 &
0.18 \\
Default configuration & 412.31 & 42.39 & 7.64 & 2.9 & 1.2 & 0.12 & 6.13
& 1.54 & 0.08 & 1.69 & 0.98 & 0.08 & 6.05 & 1.85 & 0.12 & 0.62 & 0.12 &
0.03 \\
EU stock & 416.63 & 43.96 & 4.84 & 3.02 & 1.2 & 0.11 & 6.12 & 1.51 &
0.06 & 1.76 & 1.02 & 0.14 & 6.13 & 1.93 & 0.16 & 0.60 & 0.13 & 0.05 \\
Comprehensive catalog & 415.46 & 42.65 & 4.85 & 2.91 & 1.21 & 0.13 &
6.15 & 1.51 & 0.05 & 1.73 & 0.99 & 0.1 & 6.12 & 1.86 & 0.09 & 0.62 &
0.12 & 0.03 \\
Dropped IPA & 413.22 & 42.51 & 6.82 & 2.94 & 1.19 & 0.09 & 6.07 & 1.51 &
0.08 & 1.69 & 0.97 & 0.07 & 6.07 & 1.86 & 0.12 & 0.62 & 0.12 & 0.03 \\
Dropped 3D pair embedding & 414.91 & 41.63 & 4.93 & 2.93 & 1.16 & 0.07 &
6.11 & 1.5 & 0.05 & 1.66 & 0.97 & 0.05 & 6.07 & 1.82 & 0.08 & 0.62 &
0.11 & 0.03 \\
Narrow network & 411.01 & 42.46 & 8.91 & 2.95 & 1.2 & 0.1 & 6.04 & 1.52
& 0.11 & 1.68 & 0.98 & 0.07 & 6.03 & 1.86 & 0.14 & 0.62 & 0.12 & 0.04 \\
Shallow network & 412.79 & 42.25 & 7.14 & 2.93 & 1.2 & 0.11 & 6.1 & 1.52
& 0.07 & 1.69 & 0.98 & 0.08 & 6.06 & 1.88 & 0.13 & 0.62 & 0.12 & 0.03 \\
Fast learning rate decay & 411.36 & 42.44 & 8.57 & 2.94 & 1.19 & 0.1 &
6.06 & 1.54 & 0.11 & 1.7 & 0.98 & 0.08 & 6.03 & 1.89 & 0.16 & 0.62 &
0.12 & 0.04 \\
Slow learning rate decay & 413.53 & 42.63 & 6.58 & 2.94 & 1.19 & 0.09 &
6.12 & 1.52 & 0.06 & 1.67 & 0.99 & 0.08 & 6.11 & 1.87 & 1.87 & 0.62 &
0.12 & 0.03 \\
High noise probability & 414.14 & 41.36 & 5.61 & 2.94 & 1.17 & 0.07 &
6.09 & 1.5 & 0.06 & 1.69 & 0.99 & 0.08 & 6.07 & 1.82 & 0.09 & 0.62 &
0.12 & 0.03 \\
Low noise probability & 416.82 & 45.94 & 6.43 & 2.89 & 1.26 & 0.18 & 6.3
& 1.55 & 0.19 & 1.72 & 1.03 & 0.13 & 5.91 & 1.9 & 0.26 & 0.6 & 0.14 &
0.05 \\
Large noise scale & 413.76 & 41.78 & 6.08 & 2.94 & 1.18 & 0.08 & 6.12 &
1.5 & 0.05 & 1.71 & 0.98 & 0.09 & 6.03 & 1.81 & 0.11 & 0.62 & 0.12 &
0.03 \\
Small noise scale & 413.24 & 42.9 & 6.95 & 2.93 & 1.19 & 0.1 & 6.14 &
1.53 & 0.07 & 1.7 & 0.99 & 0.09 & 6.03 & 1.86 & 0.14 & 0.62 & 0.12 &
0.04 \\
Validation set & 419.52 & 40.27 & - & 2.98 & 1.1 & - & 6.13 & 1.46 & - &
1.65 & 0.92 & - & 6.14 & 1.77 & - & 0.63 & 0.09 & - \\
Test set & 419.72 & 40.13 & - & 2.97 & 1.11 & - & 6.14 & 1.46 & - & 1.46
& 0.92 & - & 6.14 & 1.77 & - & 0.62 & 0.09 & -0.09 \\
\bottomrule()
\end{longtable}

\newpage

\hypertarget{tbl:s-prop-3d}{}
\begin{longtable}[]{@{}
  >{\raggedright\arraybackslash}p{(\columnwidth - 18\tabcolsep) * \real{0.2874}}
  >{\raggedright\arraybackslash}p{(\columnwidth - 18\tabcolsep) * \real{0.0690}}
  >{\raggedright\arraybackslash}p{(\columnwidth - 18\tabcolsep) * \real{0.0575}}
  >{\raggedright\arraybackslash}p{(\columnwidth - 18\tabcolsep) * \real{0.0575}}
  >{\raggedright\arraybackslash}p{(\columnwidth - 18\tabcolsep) * \real{0.1149}}
  >{\raggedright\arraybackslash}p{(\columnwidth - 18\tabcolsep) * \real{0.0575}}
  >{\raggedright\arraybackslash}p{(\columnwidth - 18\tabcolsep) * \real{0.0575}}
  >{\raggedright\arraybackslash}p{(\columnwidth - 18\tabcolsep) * \real{0.2069}}
  >{\raggedright\arraybackslash}p{(\columnwidth - 18\tabcolsep) * \real{0.0460}}
  >{\raggedright\arraybackslash}p{(\columnwidth - 18\tabcolsep) * \real{0.0460}}@{}}
\caption{\label{tbl:s-prop-3d}The distribution of 3D molecular
properties among molecules generated from models with different
hyperparameters (see Table \ref{tbl:s-hyperparam} for a detailed
description of each configuration). We report the mean and standard
deviation for each property, as well as the estimated Wasserstein
distance to the test set. For the first row, we report the statistics of
molecules randomly assembled from building blocks without any
drug-likeness filtering.}\tabularnewline
\toprule()
\begin{minipage}[b]{\linewidth}\raggedright
Model variant
\end{minipage} & \begin{minipage}[b]{\linewidth}\raggedright
SASA
\end{minipage} & \begin{minipage}[b]{\linewidth}\raggedright
\end{minipage} & \begin{minipage}[b]{\linewidth}\raggedright
\end{minipage} & \begin{minipage}[b]{\linewidth}\raggedright
Polar SASA
\end{minipage} & \begin{minipage}[b]{\linewidth}\raggedright
\end{minipage} & \begin{minipage}[b]{\linewidth}\raggedright
\end{minipage} & \begin{minipage}[b]{\linewidth}\raggedright
Radius of gyration
\end{minipage} & \begin{minipage}[b]{\linewidth}\raggedright
\end{minipage} & \begin{minipage}[b]{\linewidth}\raggedright
\end{minipage} \\
\midrule()
\endfirsthead
\toprule()
\begin{minipage}[b]{\linewidth}\raggedright
Model variant
\end{minipage} & \begin{minipage}[b]{\linewidth}\raggedright
SASA
\end{minipage} & \begin{minipage}[b]{\linewidth}\raggedright
\end{minipage} & \begin{minipage}[b]{\linewidth}\raggedright
\end{minipage} & \begin{minipage}[b]{\linewidth}\raggedright
Polar SASA
\end{minipage} & \begin{minipage}[b]{\linewidth}\raggedright
\end{minipage} & \begin{minipage}[b]{\linewidth}\raggedright
\end{minipage} & \begin{minipage}[b]{\linewidth}\raggedright
Radius of gyration
\end{minipage} & \begin{minipage}[b]{\linewidth}\raggedright
\end{minipage} & \begin{minipage}[b]{\linewidth}\raggedright
\end{minipage} \\
\midrule()
\endhead
& mean & std & wd & mean & std & wd & mean & std & wd \\
Randomly assembled & 667.72 & 60.84 & 41.41 & 185.77 & 58.93 & 26.24 &
4.9 & 0.71 & 0.3 \\
Default configuration & 617.86 & 53.27 & 10.07 & 162.93 & 49.92 & 2.96 &
4.56 & 0.63 & 0.05 \\
EU stock & 624.14 & 55.58 & 6.01 & 163.68 & 51.66 & 4.83 & 4.61 & 0.64 &
0.03 \\
Comprehensive catalog & 620.45 & 54.08 & 7.91 & 163.53 & 49.83 & 3.05 &
3.05 & 0.63 & 0.04 \\
Dropped IPA & 620.16 & 54.23 & 8.24 & 162.19 & 49.44 & 2.44 & 4.6 & 0.64
& 0.03 \\
Dropped 3D pair embedding & 621.13 & 52.84 & 6.8 & 161.87 & 49.19 & 2.24
& 4.59 & 0.62 & 0.02 \\
Narrow network & 618.08 & 53.9 & 10.04 & 161.95 & 49.53 & 2.56 & 4.59 &
0.64 & 0.03 \\
Shallow network & 619.66 & 54.05 & 8.62 & 162.59 & 50.19 & 3.19 & 4.6 &
0.63 & 0.02 \\
Fast learning rate decay & 616.74 & 53.97 & 11.34 & 161.7 & 49.55 & 2.63
& 4.58 & 0.64 & 0.04 \\
Slow learning rate decay & 619.5 & 53.42 & 8.54 & 162.15 & 50.4 & 3.41 &
4.57 & 0.63 & 0.05 \\
High noise probability & 620.89 & 52.67 & 6.99 & 162.3 & 49.31 & 2.3 &
4.6 & 0.63 & 0.02 \\
Low noise probability & 614.44 & 56.41 & 14.35 & 166.52 & 52.61 & 6.96 &
4.47 & 0.6 & 0.15 \\
Large noise scale & 619.44 & 52.9 & 8.45 & 162.75 & 50.11 & 3.12 & 4.59
& 0.63 & 0.03 \\
Small noise scale & 618.84 & 54.05 & 9.38 & 162.16 & 49.65 & 2.65 & 4.57
& 0.63 & 0.05 \\
Validation set & 627.61 & 50.87 & - & 162.5 & 47.12 & - & 4.62 & 0.62 &
- \\
Test set & 627.52 & 50.49 & - & 162.57 & 46.89 & - & 4.61 & 0.61 & - \\
\bottomrule()
\end{longtable}

\newpage

\hypertarget{tbl:s-mmd}{}
\begin{longtable}[]{@{}
  >{\raggedright\arraybackslash}p{(\columnwidth - 10\tabcolsep) * \real{0.2747}}
  >{\raggedright\arraybackslash}p{(\columnwidth - 10\tabcolsep) * \real{0.1429}}
  >{\raggedright\arraybackslash}p{(\columnwidth - 10\tabcolsep) * \real{0.1429}}
  >{\raggedright\arraybackslash}p{(\columnwidth - 10\tabcolsep) * \real{0.1319}}
  >{\raggedright\arraybackslash}p{(\columnwidth - 10\tabcolsep) * \real{0.1319}}
  >{\raggedright\arraybackslash}p{(\columnwidth - 10\tabcolsep) * \real{0.1758}}@{}}
\caption{\label{tbl:s-mmd}Quantitative measurement of the quality of
generated samples. The first two columns indicate the sample diversity
measured using Tanimoto and USRCAT fingerprints. The third and fourth
column shows the 2D and 3D MMD values, which measures the ability of the
network to correctly model the distribution in the chemical space. The
final column contains the average RMSD after conformers are relaxed
using MMFF94s forcefield. Each row represents a different hyperparameter
configuration, as detailed in Table \ref{tbl:s-hyperparam}. The first
row represents molecules randomly assembled from building blocks without
any drug-likeness filtering.}\tabularnewline
\toprule()
\begin{minipage}[b]{\linewidth}\raggedright
Model variant
\end{minipage} & \begin{minipage}[b]{\linewidth}\raggedright
Diversity(2D)
\end{minipage} & \begin{minipage}[b]{\linewidth}\raggedright
Diversity(3D)
\end{minipage} & \begin{minipage}[b]{\linewidth}\raggedright
MMD(2D)
\end{minipage} & \begin{minipage}[b]{\linewidth}\raggedright
MMD(3D)
\end{minipage} & \begin{minipage}[b]{\linewidth}\raggedright
Mean RMSD(\(\text{\AA}\))
\end{minipage} \\
\midrule()
\endfirsthead
\toprule()
\begin{minipage}[b]{\linewidth}\raggedright
Model variant
\end{minipage} & \begin{minipage}[b]{\linewidth}\raggedright
Diversity(2D)
\end{minipage} & \begin{minipage}[b]{\linewidth}\raggedright
Diversity(3D)
\end{minipage} & \begin{minipage}[b]{\linewidth}\raggedright
MMD(2D)
\end{minipage} & \begin{minipage}[b]{\linewidth}\raggedright
MMD(3D)
\end{minipage} & \begin{minipage}[b]{\linewidth}\raggedright
Mean RMSD(\(\text{\AA}\))
\end{minipage} \\
\midrule()
\endhead
Randomly assembled & 0.120 & 0.156 & 0.003604 & 0.003492 & 1.142
(ETKDG) \\
Default configuration & 0.121 & 0.162 & 0.000158 & 0.000156 & 0.692 \\
EU stock & \textbf{0.123} & 0.161 & 0.000798 & 0.000109 & 0.725 \\
Comprehensive catalog & 0.121 & 0.162 & 0.000134 & 0.000122 & 0.691 \\
Dropped IPA & 0.121 & 0.161 & 0.000195 & 0.000068 & 0.696 \\
Dropped 3D pair embedding & 0.122 & 0.162 & \textbf{0.000082} &
\textbf{0.000045} & 0.718 \\
Narrow network & 0.121 & 0.161 & 0.000256 & 0.000066 & 0.716 \\
Shallow network & 0.121 & 0.161 & 0.000204 & 0.000059 & 0.707 \\
Fast learning rate decay & 0.121 & 0.161 & 0.000224 & 0.000126 &
0.734 \\
Slow learning rate decay & 0.121 & 0.162 & 0.000183 & 0.000120 &
\textbf{0.672} \\
High noise probability & 0.122 & 0.162 & 0.000171 & 0.000060 & 0.685 \\
Low noise probability & 0.120 & \textbf{0.165} & 0.000475 & 0.000954 &
0.795 \\
Large noise scale & 0.122 & 0.162 & 0.000163 & 0.000080 & 0.685 \\
Small noise scale & 0.121 & 0.162 & 0.000212 & 0.000143 & 0.713 \\
\bottomrule()
\end{longtable}

\newpage
\end{landscape}
}